  \documentclass[sn-mathphys-num]{sn-jnl}% Math and Physical Sciences Numbered Reference Style 

  \usepackage{bera}
  \usepackage{graphicx}%
  \usepackage{multirow}%
  \usepackage{amsmath,amssymb,amsfonts}%
  \usepackage{amsthm}%
  \usepackage{fix-cm}          % 允许自由缩放字体
  \usepackage{type1cm}         % 让 Type1 字体可以缩放，避免缺号
  \usepackage{newtxtext,newtxmath}  % 提供完整数学字体，包括 \mathscr
  \usepackage[title]{appendix}%
  \usepackage{xcolor}%
  \usepackage{textcomp}%
  \usepackage{manyfoot}%
  \usepackage{booktabs}%
  \usepackage{algorithm}%
  \usepackage{algorithmicx}%
  \usepackage{algpseudocode}%
  \usepackage{listings}%
  \usepackage{makecell}
  \usepackage{tcolorbox}
  \usepackage{array}
  \usepackage{booktabs}
  \theoremstyle{thmstyleone}%

  \theoremstyle{thmstyletwo}%

  \theoremstyle{thmstylethree}%

  \newboolean{showcomments}
  \setboolean{showcomments}{true} % toggle to show or hide comments
  \ifthenelse{\boolean{showcomments}}
  {\newcommand{\nb}[2]{
      \fcolorbox{black}{yellow}{\bfseries\sffamily\scriptsize#1}
      {\sf\small$\blacktriangleright$\textit{#2}$\blacktriangleleft$}
    }
    
  }
  {\newcommand{\nb}[2]{}
    
  }

  \newcommand{\methodname}{\textsc{SAGE}}
  
  \colorlet{punct}{red!60!black}
  \definecolor{background}{HTML}{EEEEEE}
  \definecolor{delim}{RGB}{20,105,176}
  \colorlet{numb}{magenta!60!black}

  \lstdefinelanguage{json}{
      basicstyle=\normalfont\footnotesize,
      numbers=left,
      numberstyle=\scriptsize,
      stepnumber=1,
      numbersep=8pt,
      showstringspaces=false,
      breaklines=true,
      frame=lines,
      backgroundcolor=\color{background},
      literate=
      *{0}{{{\color{numb}0}}}{1}
        {1}{{{\color{numb}1}}}{1}
        {2}{{{\color{numb}2}}}{1}
        {3}{{{\color{numb}3}}}{1}
        {4}{{{\color{numb}4}}}{1}
        {5}{{{\color{numb}5}}}{1}
        {6}{{{\color{numb}6}}}{1}
        {7}{{{\color{numb}7}}}{1}
        {8}{{{\color{numb}8}}}{1}
        {9}{{{\color{numb}9}}}{1}
        {:}{{{\color{punct}{:}}}}{1}
        {,}{{{\color{punct}{,}}}}{1}
        {\{}{{{\color{delim}{\{}}}}{1}
        {\}}{{{\color{delim}{\}}}}}{1}
        {[}{{{\color{delim}{[}}}}{1}
        {]}{{{\color{delim}{]}}}}{1},
  }

\begin{document}

  \title[Article Title]{SAGE: Semantic-Aware Gray-Box Game Regression Testing with Large Language Models}

  % \author*[1,2]{\fnm{First} \sur{Author}}\email{iauthor@gmail.com}

  \author[1]{\fnm{Jinyu} \sur{Cai}}\email{bluelink@toki.waseda.jp}
  \author[1]{\fnm{Jialong} \sur{Li}}\email{lijialong@fuji.waseda.jp}
  \author[2]{\fnm{Nianyu} \sur{Li}}\email{li\_nianyu@pku.edu.cn}
  \author[3]{\fnm{Zhenyu} \sur{Mao}}\email{zhenyumao2-c@my.cityu.edu.hk}
  \author*[4]{\fnm{Mingyue} \sur{Zhang}}\email{myzhangswu@swu.edu.cn}
  \author[5]{\fnm{Kenji} \sur{Tei}}\email{tei@comp.isct.ac.jp}

  \affil[1]{\orgname{Waseda University}, \orgaddress{\city{Tokyo}, \country{Japan}}}
  \affil[2]{\orgname{Independent Researcher}, \orgaddress{\city{Beijing}, \country{China}}}
  % \affil[3]{\orgname{Dalian Maritime University}, \orgaddress{\city{Dalian}, \country{China}}}
  \affil[3]{\orgname{City University of Hong Kong}, \orgaddress{\city{Hong Kong}, \country{China}}}
  \affil[4]{\orgname{Southwest University}, \orgaddress*{\city{Chongqing}, \country{China}}}
  \affil[5]{\orgname{Institute of Science Tokyo}, \orgaddress{\city{Tokyo}, \country{Japan}}}

  \abstract{
  The rapid iteration cycles of modern live-service games make regression testing indispensable for maintaining quality and stability. However, existing regression testing approaches face critical limitations, especially in common gray-box settings where full source code access is unavailable: they heavily rely on manual effort for test case construction, struggle to maintain growing suites plagued by redundancy, and lack efficient mechanisms for prioritizing relevant tests. These challenges result in excessive testing costs, limited automation, and insufficient bug detection.
  To address these issues, we propose \methodname{}, a semantic-aware regression testing framework for gray-box game environments. \methodname{} systematically addresses the core challenges of test generation, maintenance, and selection. It employs LLM-guided reinforcement learning for efficient, goal-oriented exploration to automatically generate a diverse foundational test suite. Subsequently, it applies a semantic-based multi-objective optimization to refine this suite into a compact, high-value subset by balancing cost, coverage, and rarity. Finally, it leverages LLM-based semantic analysis of update logs to prioritize test cases most relevant to version changes, enabling efficient adaptation across iterations.
  We evaluate \methodname{} on two representative environments, Overcooked Plus and Minecraft, comparing against both automated baselines and human-recorded test cases. Across all environments, \methodname{} achieves superior bug detection with significantly lower execution cost, while demonstrating strong adaptability to version updates.  
  }

  \maketitle

  \section{Introduction}
  In recent years, digital games have evolved into large-scale, continuously updated software systems.
  Under the ``game-as-a-service'' model, developers frequently release content updates—such as new quests, gameplay elements, and balance adjustments—to continuously evolve the game system~\cite{newzoo2024}. However, each update inevitably modifies part of the game logic or data, introducing potential inconsistencies or regressions that may compromise existing functionalities, degrade user experience, or cause economic loss~\cite{RLReg}.
  Because modern games typically involve complex state spaces, real-time interactions, and tightly coupled subsystems, ensuring the functional stability of updates within a fast-paced iterative workflow has become a major technical challenge in game development.

  Within this context, regression testing is widely regarded as a key technique to ensure game quality. By re-executing existing test cases, regression testing verifies whether new versions preserve previously implemented functionalities~\cite{agrawal1993incremental}. However, given the enormous state spaces and high uncertainty of game environments, existing regression testing methods are still highly dependent on manual execution, consuming both time and human resources. It is reported that regression testing can account for more than 80\% of the total testing cost in real projects~\cite{gligoric2015practical}. As a result, many companies only conduct full-scale regression testing during major version releases, leaving potential quality risks in frequent minor updates. Consequently, the development of efficient and automated game regression testing methods has become an urgent research direction.

  Early regression testing relied heavily on human testers to manually execute and verify tasks. While this approach ensured reliability, it imposed prohibitive labor costs. To alleviate this issue, researchers proposed record-and-replay mechanisms~\cite{netravali2015mahimahi}, which capture real player operations and transform them into reusable test cases, thereby reducing repetitive effort. Later, Ostrowski et al.~\cite{ostrowski2013automated} developed a GUI-based recording framework that simplified test case construction, enabling testers without programming expertise to produce high-quality scripts. With the advent of machine learning, reinforcement learning (RL) methods were introduced into game testing~\cite{9619048}. Unlike fixed action sequences, RL agents learn behavioral strategies by dynamically interacting with the environment, with the learned policy itself serving as a test case. This allows RL agents to actively explore and cover a broader set of potential scenarios. More recently, the white-box GameRTS method~\cite{GameRTS} advanced the systematization and efficiency of regression testing by introducing a regression test selection (RTS) workflow for games. By analyzing source code and resources, the method can intelligently identify the most relevant test cases to re-execute, effectively avoiding unnecessary redundancy while maintaining a high bug detection rate.

  However, testing teams, especially in large game development teams, are often independent or outsourced from the development team, and operate in gray-box settings. This setting represents a realistic condition: while the source code and internal implementation details are inaccessible (unlike white-box testing), testers can still access and observe structured runtime information, such as game APIs, event callbacks, runtime logs (e.g., object states, player actions), and high-level developer documentation (e.g., changelogs).
  Under this practical constraint of the gray-box setting, how to achieve efficient and automated regression testing methods remains a challenge. Specifically, this can be subdivided into the following three problems: 
  (i) the Foundation Issue (test suite construction): Regression testing often presupposes the existence of a high-quality, high-coverage test suite. However, in complex gray-box game projects, constructing this initial, high-quality suite is itself a significant challenge. Specifically, the existing RL-based exploration suffers from low efficiency and local optima, while traditional record-and-replay methods depend heavily on human effort and lack scalability.
  (ii) the Maintenance Issue (test suite maintenance): As game versions are continuously updated, the foundational test suite continuously grows in size, and it becomes progressively filled with outdated, invalid, and redundant test cases. Therefore, the second challenge is how to maintain a compact, high-value static test suite that maximizes regression detection while minimizing wasteful execution.
  (iii) the Selection Issue (test suite selection): Given a high-quality, well-maintained general-purpose suite, the rapid update frequency often makes it unrealistic to execute the entire suite within limited time constraints. Therefore, the third challenge is: given a specific version update and its associated documentation, how can we dynamically identify and prioritize the subset of test cases from the test suite that are most relevant to the changes, to maximize the bug-detection rate within a constrained testing window.

  While gray-box testing restricts direct access to source code, it exposes rich, human-understandable semantic artifacts—runtime logs, event traces, and changelogs—that reflect a game’s underlying logic and behavioral dynamics. Modern large language models (LLMs) excel at semantic understanding and reasoning; they can interpret these artifacts and ground them into machine-executable actions, enabling systematic regression testing without internal code analysis. We therefore propose \methodname{}, a semantic-aware regression testing framework for gray-box game environments. At its core, \methodname{} installs the LLM as a semantic orchestrator, transforming heuristic practices into a systematic, data-driven process that operates entirely on log-level information. 
  To realize this vision, \methodname{} systematically resolves the three aforementioned issue through three interconnected mechanisms: (1) for the Foundation Issue, a semantic-driven generation mechanism where an LLM guides a reinforcement learning (RL) agent to conduct efficient, goal-oriented exploration and automatically assemble a diverse foundational test suite;  
  (2) for the Maintenance Issue, a semantics-informed multi-objective optimization process that balances cost, coverage, and behavioral rarity, refining the expanding repository into a compact, high-value static subset; 
  (3) for the Selection Issue, an LLM-based update interpreter that reads natural-language change logs in the gray-box setting and translates version deltas into test priorities, dynamically directing resources toward the most relevant test cases.

  Our contributions are as follows:
  \begin{itemize}
      \item We design \methodname{}, an end-to-end unified gray-box regression testing framework. It automates test case generation, multi-objective optimization, and update-aware test case prioritization into a closed-loop, semantic-driven process, systematically addressing the core challenges in gray-box game testing scenarios.
      \item We propose a test case optimization and selection method that uses semantics as a bridge. The core of this method lies in its use of LLMs to translate high-level, unstructured information (e.g., abstract test goals, natural language update logs) into structured, machine-executable testing strategies (e.g., quantifiable optimization metrics, precise test priorities), enabling intelligent test decisions in environments without source code.
      \item We conduct a comprehensive empirical study in two complex game environments—Overcooked Plus and Minecraft. The results demonstrate that \methodname{} significantly outperforms existing baselines in terms of coverage, efficiency, and bug detection. To promote reproducibility in the field, we are open-sourcing the replication suite used in our experiments.
  \end{itemize}

  The rest of this paper is organized as follows. Section~\ref{sec:relatedwork} reviews related work. Section~\ref{sec:proposal} presents \methodname{} in detail. Section~\ref{sec:evaluation} describes our experimental setup and discusses the results. Finally, Section~\ref{sec:conclusion} concludes the paper and outlines future research directions.

  \section{Related Work}
  \label{sec:relatedwork}
  Game testing has always played a critical role in the software development lifecycle of games, aiming to ensure stability, completeness, and user experience during both initial release and subsequent updates~\cite{DBLP:journals/sbcjis/DuarteMDNE24,DBLP:conf/dsa/ZhangZAI23}. With the increasing complexity of modern games and the prevalence of the ``game-as-a-service'' model, testing approaches have gradually evolved from manual validation to more automated and intelligent solutions. In this section, we review related work from three perspectives: traditional regression testing methods, agent-based automated testing methods, and the applications of LLMs in game agents and testing.

  \subsection{Traditional Regression Testing Approaches}

  In the early stages of game testing, the most common practice was to rely on human testers to manually execute and validate tasks\cite{politowski2021survey}. While this approach ensured reliability, it was highly inefficient and did not scale to the vast state spaces of modern games.
  
  To reduce the overhead of repetitive work, prior work proposed \emph{record-and-replay} mechanisms~\cite{ostrowski2013automated}. These methods capture real player interactions and transform them into reusable regression test cases, thereby reducing manual repetition. However, such approaches still relied on testers to identify key interactions, and the recorded traces often contained redundant operations, limiting overall efficiency.

  Later, \emph{script-based testing methods} emerged, where testers designed macros or control scripts to simulate player behavior~\cite{spronck2006adaptive,mioto2025mapping}. These methods worked well in fixed scenarios and quests, but script design often required deep domain knowledge of both gameplay mechanics and testing principles. As games became more complex, the cost of maintaining and extending scripts grew rapidly.

  Building on this, more advanced forms of automation have also been explored. For instance, the GameRTS method~\cite{GameRTS} constructs game state transition graphs through static code analysis and applies RTS techniques to identify affected test cases, thus improving efficiency while reducing maintenance overhead. However, such methods typically assume full access to source code and internal resources—an assumption that is often unrealistic in practice, where testing is usually performed under limited code accessibility~\cite{politowski2021survey}.

  \subsection{Agent-Based Automated Testing Approaches}

  Compared to script-driven testing, which often struggles with scalability and adaptability in dynamic environments, agent-based exploration methods~\cite{stahlke2020artificial,iftikhar2015automated,stahlke2019artificial} employ autonomous agents to simulate diverse player behaviors. 

  Early-stage works have explored a variety of approaches, including model-based testing for platformers~\cite{iftikhar2015automated} and behavior-tree agents for playability analysis and level evaluation~\cite{stahlke2019artificial,stahlke2020artificial}, offering enhanced automation and better support for early-stage design and gameplay validation. 

  Later, reinforcement learning (RL) was introduced into agent-based approaches and has attracted significant attention, as RL agents learn behavioral policies through trial-and-error interactions with the environment, enabling broader coverage and greater adaptability. 
  For example, curiosity-driven intrinsic rewards have been introduced to guide agents toward rarely visited states, improving behavioral diversity and bug detection~\cite{9619048}. 
  Prior work has also developed RL-based frameworks for version-aware regression testing~\cite{RLReg}. In particular, one such framework compares agent behavior trajectories across game versions to shift the focus from pure exploration to version-aware correctness. It incorporates differences in observed states under identical task conditions into the reward function, enabling the agent to detect potential regressions introduced by updates.
  The Wuji framework~\cite{zheng2019wuji} combines deep RL with evolutionary strategies to balance task completion and exploration, further enhancing coverage. Other studies have also attempted to expand exploration capabilities using evolutionary algorithms or preference-driven agents~\cite{guerrero-romero_using_2018}.

  Nevertheless, these methods primarily focus on discovering new states rather than validating the correctness of updates, and thus lack regression-oriented specificity. Consequently, they often produce large amounts of redundant actions, increasing testing cost and execution overhead. Moreover, RL models require frequent retraining across versions, leading to prohibitive computational expenses in iterative development workflows.

  \subsection{Large Language Models in Software Testing and Games}
  LLMs have recently emerged as a transformative force in software testing, enabling a shift from manually designed heuristics to semantics-driven automation~\cite{wang2024software}. 
  By leveraging deep semantic understanding, reasoning, and multi-step planning capabilities, LLMs can process unstructured artifacts and generate executable test assets, thereby redefining automation across the entire software testing lifecycle.
  Wang et al.~\cite{wang2024software} refer to this emerging line of work collectively as LLM4TEST, which spans the role of LLMs throughout the software testing lifecycle. 

  \textit{Unit Test Case Generation.}
  Unit test case generation automates the creation of test cases for verifying individual functions or methods, serving as the foundation of automated testing. Representative studies fine-tuned pre-trained transformers on Java corpora to translate focal methods into tests~\cite{Tufano2020}, enhanced generation with assertion-based prompts in A3Test~\cite{Alagarsamy2023}, proposed adaptive prompts for ChatGPT in ChatUniTest~\cite{Xie2023}, and integrated mutation testing feedback to refine generated cases~\cite{Dakhel2023}.

  \textit{Test Oracle Generation.}
  Test oracle generation focuses on producing expected outputs or assertions that verify correctness in generated tests. Representative work fine-tuned T5 for assertion synthesis~\cite{Mastropaolo2023}, demonstrated dual-language pre-training to improve oracle precision~\cite{TufanoAsserts2022}, and applied retrieval-augmented prompting to extract semantically similar examples, achieving higher oracle accuracy without requiring fine-tuning~\cite{Nashid2023}.

  \textit{System-Level Test Input Generation.}
  System-level input generation targets the testing of interactive systems, where LLMs are applied to generate semantic actions or valid inputs. Existing systems include QTypist for context-aware GUI input generation~\cite{LiuFill2022}, GPTDroid for dialogue-based mobile testing~\cite{LiuTestingExpert2023}, and LLM-based fuzzing approaches for exposing edge cases in deep-learning frameworks~\cite{DengEdgeCase2023,DengZeroShot2022}.

  \textit{Bug Analysis and Debugging.}
  Bug analysis and debugging involve using LLMs to interpret, localize, and reproduce software bugs. Representative examples include iTiger for automatic bug title generation~\cite{ZhangITiger2022}, Detect–Localize–Repair for bug localization and fixing with CodeT5~\cite{BuiDLR2022}, self-debugging mechanisms in which LLMs correct their outputs using runtime feedback~\cite{ChenSelfDebug2023}, and prompting-based replay of Android crashes~\cite{FengAdbGPT2023}.

  \textit{Program Repair.}
  Program repair applies LLMs to automatically generate or refine code patches for defective software. Prior studies have used GPT-based repair for JavaScript~\cite{Lajko2022} and leveraged retrieval-based bug–fix examples to improve patch accuracy~\cite{WangExamples2023}.

  In summary, the natural language understanding capabilities of LLMs enable more intelligent automation by integrating richer contextual information into the testing workflow. Inspired by this trend, and adapting to prevalent gray-box settings in game testing, our work leverages game-specific context and uses semantic information to connect the full lifecycle of test case generation, optimization, and selection.

  \subsection{Position of This Study}
  In this study, we position our work within a gray-box testing scenario—a realistic middle ground between white-box and black-box settings. In such scenarios, the source code and internal implementation details are inaccessible, yet the system typically exposes a limited degree of structured interaction interfaces, such as APIs, event callbacks, and runtime logs. Testers can observe runtime information (e.g., object states, player actions, and event traces) but cannot directly analyze source-level artifacts such as code differences or class hierarchies. This setting aligns with practice: in commercial game development, testing teams are often organizationally separate from development teams. Testers usually have access to certain high-level APIs and detailed runtime logs for testing purposes, while underlying scripts and resource files remain closed due to intellectual property protection and modular development workflows~\cite{politowski2021survey,ixie2024gametesting}.

  Under this premise, despite notable progress, several limitations remain in existing studies:
  Traditional regression testing methods have reduced manual effort but still suffer from the high cost of script maintenance and dependence on source code, which limits their applicability.
  Agent-based approaches—particularly RL agents—offer strong automation and coverage capabilities but mainly emphasize broad exploration rather than update validation, resulting in redundant actions and high retraining costs.
  The emergence of LLMs brings new opportunities, as they excel at handling semantic and unstructured information and can further enhance automation; however, their potential in regression testing has not yet been fully explored.

  Based on these observations, we propose \methodname{}, a unified gray-box regression testing framework that combines the semantic reasoning capabilities of LLMs with graph-based regression testing. The framework eliminates the reliance on manual recording and script-based testing by enabling semantic-driven exploration for efficient and goal-oriented test case generation. It further enhances test case quality through a multi-objective optimization process that balances coverage, cost, and behavioral diversity, and achieves adaptive regression validation via an Update-Aware Test Case Prioritization mechanism that leverages semantic information extracted from natural-language update logs. Notably, \methodname{} operates solely on log-level information—such as player actions, in-game event traces, and system runtime logs—without requiring any access to source code or internal assets. This design allows the framework to achieve end-to-end automation, high scalability, and semantic adaptability, aligning with the practical constraints of industrial gray-box testing environments.
  % In doing so, our work not only continues the trajectory of automation in game testing but also provides a systematic, scalable, and efficient solution for regression testing in high-frequency update environments.

  \section{Proposal}
  \label{sec:proposal} 

  \subsection{Overview}
  \begin{figure}[h!]
      \centering
      \includegraphics[width=0.99\linewidth]{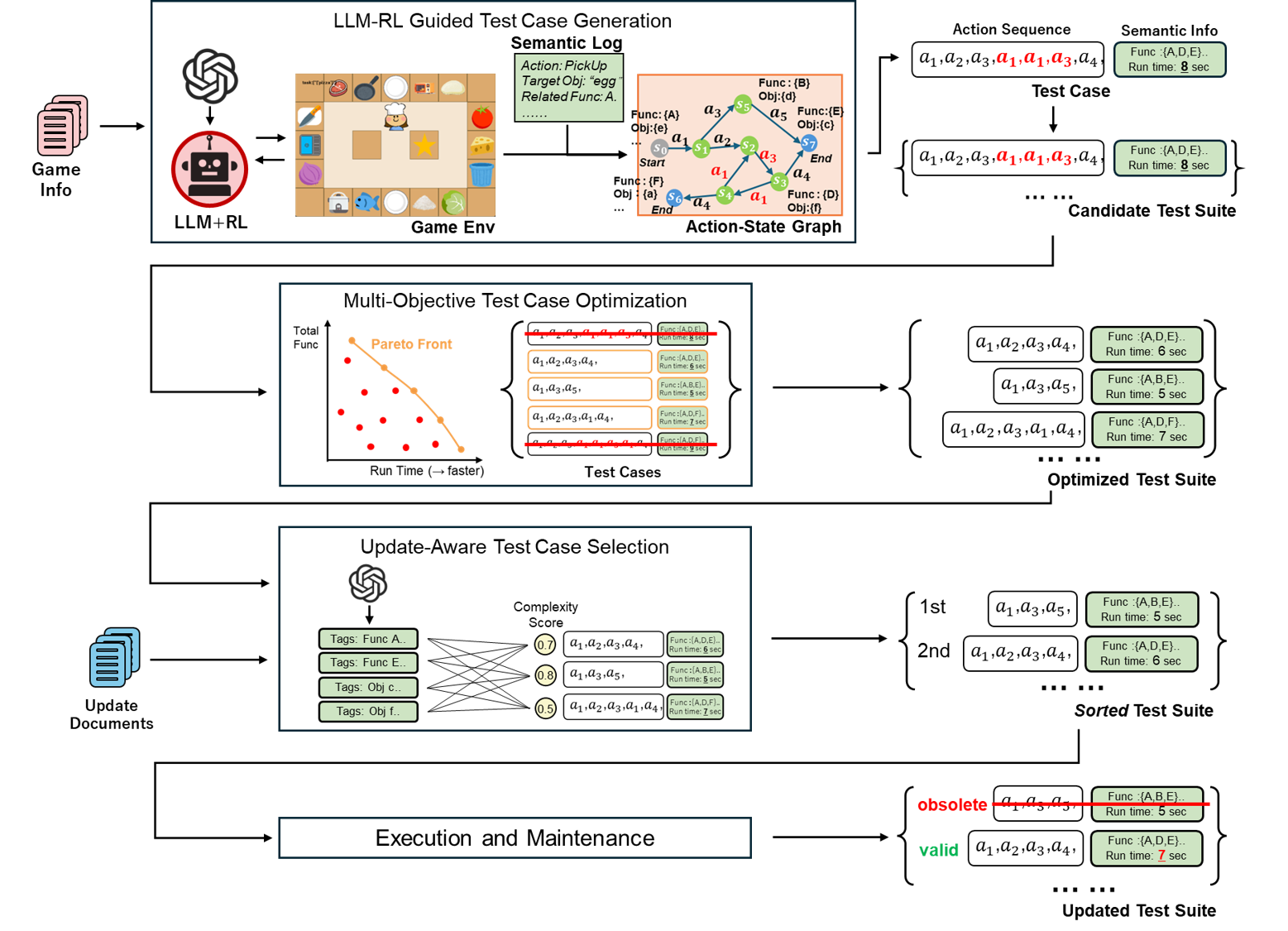}
      \caption{Overview of \methodname{}. The framework consists of four main stages: (1) LLM-RL guided test case generation, (2) multi-objective test case optimization, (3) update-aware test case selection, and (4) test case execution and maintenance. These stages together enable automated, scalable, and update-sensitive regression testing in complex game environments. For better visibility and easier understanding, the state information ($s$) contained in test cases is omitted in this figure.}   
      \label{fig:overview}
  \end{figure}

  This section introduces \methodname{}, a unified gray-box regression testing framework. As illustrated in Figure~\ref{fig:overview}, \methodname{} operates as a semantic-driven closed-loop system where each stage logically builds upon the output of the last, progressively refining test suite from raw exploration data into a prioritized execution list.

  \begin{itemize}
    \item \textcolor{black}{The process begins with LLM-RL Guided Test Case Generation, where LLM-generated seed trajectories and RL exploration are combined to build a state--action transition graph of task-relevant behaviors.}
    \item \textcolor{black}{The framework then applies Multi-Objective Test Case Optimization to evaluate graph paths by cost, coverage, and behavioral rarity, retaining a compact Pareto-optimal subset as the reusable regression test suite.}
    \item \textcolor{black}{For a specific version update, Update-Aware Test Case Selection extracts semantic tags from the update log and ranks the optimized test cases by their relevance to the update and their functional complexity.}
    \item \textcolor{black}{Finally, Test Case Execution and Maintenance re-executes the prioritized cases on the new version and uses the outcomes to refresh the metadata of the underlying test repository.}
  \end{itemize}

  \subsection{LLM-RL Guided Test Case Generation}
  \textcolor{black}{As the starting point of the \methodname{} pipeline, this stage initializes the foundational test assets that capture exploratory behaviors and environment interactions.
  These assets serve as the raw material for deriving executable test cases.
  To enable efficient and scalable generation, the stage integrates LLMs, RL, and graph-based modeling: LLMs provide task-oriented action priors, RL expands and diversifies the behavioral space, and the resulting transitions are aggregated into a unified state--action transition graph.}

  \subsubsection{Seed Trajectory Generation with LLMs}
  \textcolor{black}{This module begins by prompting an LLM to generate seed trajectories that can accomplish a given task. The core idea is to expose the current game situation to the LLM as a structured state representation that preserves task-relevant information while abstracting away low-level runtime details, so that planning can be performed at the level of gameplay intent rather than raw system state. Each prompt therefore specifies the environment context, the current structured state representation, the task objective, and the available actions; when available, previously successful solutions can also be included to encourage strategy diversity. Based on this information, the LLM produces a task-oriented sequence of high-level executable actions, which is then encoded into transition tuples \((s_t, a_t, s_{t+1})\), where \(s_t\) and \(s_{t+1}\) denote game states and \(a_t\) is the executed action. These transitions form the initial structure of the state--action transition graph, outlining the behavioral space surrounding the task.}

  \textcolor{black}{Representative prompt examples for both Overcooked Plus and Minecraft are provided in Appendix~\ref{appendix:llm_prompts} (Listings~1 and~2).}

  \subsubsection{Policy Learning and Guided Exploration}
  \label{sec:RLtrain}
  Although LLM-generated trajectories are generally task-effective, they are limited in behavioral diversity and expensive to obtain. To enhance state space coverage, we introduce RL agents to perform environment-driven exploration. Specifically, we use behavior cloning on the LLM-generated seed trajectories to train an initial policy \(\pi_0(a \mid s)\), which serves as a strong prior for the RL agent and avoids inefficient exploration from random initialization.

  Subsequently, the agent explores the environment using a composite reward function:
  \[
  r(s_t,a_t) = r_{\text{goal}}(s_t) + \frac{1}{n(s_t)},
  \]
  where \(r_{\text{goal}}(s_t)\) denotes the terminal reward for reaching a goal state, \(n(s_t)\) is the number of times state \(s_t\) has been visited. This formulation encourages the agent to prioritize novel states while still completing the task, thereby increasing trajectory diversity.

  \subsubsection{Graph Construction and Test Case Derivation}

  \textcolor{black}{All observed transitions during exploration are aggregated into a state--action transition graph \(\mathcal{G} = (S, A, E)\), where:}
  \begin{itemize}
      \item \textcolor{black}{\(S = \{s_0, s_1, \dots\}\): the set of unique abstract states observed during exploration. State abstraction maps low-level observations to compact, task-relevant variables that preserve the semantic information needed for planning and test reasoning.}
      \item \textcolor{black}{\(A = \{a_0, a_1, \dots\}\): the set of abstract actions available to the agent. Action abstraction maps low-level control signals to semantically meaningful executable operations, including parameterized actions when needed.}
      \item \textcolor{black}{\(E \subseteq S \times A \times S\): the set of directed transitions, where \(e = (s, a, s')\) records that executing action \(a\) in state \(s\) leads to state \(s'\). Each edge may additionally store metadata used for downstream optimization and prioritization, such as execution cost and interaction descriptors (see Section~\ref{sec:method_metric}).}
  \end{itemize}

  A candidate test case is operationalized as a path \(\tau\) from the initial state \(s_0\) to a task-completion terminal state \(s_t \in S_{\text{term}}\):
  \[
  \tau = (s_0 \xrightarrow{a_0} s_1 \xrightarrow{a_1} \dots \xrightarrow{a_{n-1}} s_n), \quad \text{where } s_n \in S_{\text{term}}.
  \]
  \textcolor{black}{Here, \(S_{\text{term}}\) denotes the set of states that satisfy the task-completion criterion for the current environment, either by reaching an explicit goal state or by receiving an environment-provided success signal. The graph itself may contain revisits and cycles, but candidate-path extraction uses bounded search depth and a cap on the number of extracted paths to keep enumeration tractable in large state spaces.}

  These candidate paths form the input to the multi-objective optimization and update-aware selection modules.

  \subsection{Multi-Objective Test Case Optimization}
  \textcolor{black}{In this stage, we compress the large number of candidate test cases generated from the state--action graph based on their associated semantic information, producing a subset that is both representative and execution-efficient. To avoid the information loss caused by collapsing test cases into a single scoring function, each test case is encoded as a multi-dimensional feature vector based on different semantic metrics. We then apply a multi-objective optimization approach, selecting the set of non-dominated paths on the Pareto front to retain diverse behavioral semantics while eliminating redundant cases.}

  \subsubsection{Objective Formulation and Metric Definition}
  \label{sec:method_metric}

  Inspired by prior work on multi-objective test case optimization~\cite{de2014hybrid,mondal2015exploring,de2011multi}, and adapted to the specific requirements of game testing under limited code accessibility, we define three categories of objectives: cost, coverage, and rarity. Path length and execution time reflect resource cost; coverage metrics capture test value; and we introduce a novel n-gram rarity metric to prioritize paths with potentially high bug triggering capability.

  Given a candidate path \(\tau_i\), we define:

  \textbf{(1) Cost-related Objectives}:
  \begin{itemize}
      \item \emph{Path Length} \(T_i\): number of actions in the path;
      \item \emph{Execution Time} \(R_i\): estimated time required to execute the path.
  \end{itemize}

  \textbf{(2) Coverage-related Objectives}:
  \begin{itemize}
      \item \emph{State Coverage} \(C_i\): number of unique states visited;
      \item \emph{Action Diversity} \(A_i\): number of distinct actions in the path;
      \item \emph{Object Diversity} \(O_i\): number of distinct in-game objects interacted with;
      \item \emph{Scene Coverage} \(S_i\): number of unique map regions visited;
      \item \emph{UI Component Coverage} \(U_i\): number of distinct UI components triggered.
  \end{itemize}

  \textbf{(3) Rarity Objective}:
  \begin{itemize}
      \item \emph{n-gram Rarity} \(N_i\): rarity score of local action subsequences based on their global frequency across all paths.
  \end{itemize}

  These features are encoded as a vector for each path and used in multi-objective selection.

  \subsubsection{Modeling Behavioral Rarity}

  \textcolor{black}{In real-world game testing, bugs are often triggered by specific local combinations of states and actions rather than by global path patterns. For instance, transitions such as "switching weapons at the exact moment of skill-cooldown completion" or "picking up an item when the inventory is almost full" are typical failure-inducing scenarios.}

  \textcolor{black}{To capture such critical behaviors, we define the n-gram rarity \(N_i\) by computing the inverse frequency of short action subsequences across the candidate path pool. In the reported experiments, this general formulation is instantiated with 2-gram subsequences.} This metric favors paths with rare yet plausible behaviors, thereby increasing the likelihood of exposing edge-case bugs without inflating the optimization problem's dimensionality.

  \subsubsection{Pareto-Optimal Path Selection}

  Since the objectives are often in conflict (e.g., longer paths cover more but cost more), we adopt a Pareto front-based approach to select test cases. 
  A path \(\tau_i\) is Pareto-optimal if there is no other path \(\tau_j\) such that:
  \[
  \forall k, \quad f_k(\tau_j) \leq f_k(\tau_i) \quad \text{and} \quad \exists k, f_k(\tau_j) < f_k(\tau_i),
  \]
  where \(f_k\) denotes the \(k\)-th objective (e.g., cost, coverage, or rarity).

  The final optimized test suite \(\mathcal{P}^*\) is composed of all Pareto-optimal paths, representing an effective balance among competing objectives.

  \subsection{Update-Aware Test Case Selection}
  \textcolor{black}{This stage prioritizes test cases according to their semantic relevance to an update and their functional complexity. We use an LLM to summarize each update log into a concise set of semantic tags describing affected gameplay elements, and compare these tags with the semantic metadata attached to each test case.} A complexity score is calculated based on interaction diversity, and both signals are combined to produce update-aware execution priorities.

  \subsubsection{Semantic Change Extraction from Update Logs}

  Given an update document \(U\) (e.g., update logs), we use an LLM to extract a set of structured semantic tags:
  \[
  \mathcal{K} = \{k_1, k_2, \dots, k_n\},
  \]
  \textcolor{black}{where each \(k_i\) denotes a concise descriptor of an affected gameplay element, such as an item, action, mechanic, scene, or UI component. The prompt constrains the model to return salient gameplay-related descriptors rather than raw log text, and representative prompts are provided in Appendix~\ref{appendix:llm_prompts}.} These tags act as lightweight proxies for functional changes and help localize affected behaviors without requiring code access.

  \subsubsection{Test Case Prioritization}

  Each test case \(t_j \in \mathcal{T}\) is annotated with semantic metadata (actions, objects, scenes, UI components, and states) and encoded as a vector \(\mathbf{v}_{t_j}\), using a pre-trained Sentence-BERT model~\cite{reimers-gurevych-2019-sentence}. \textcolor{black}{Each extracted tag \(k_i \in \mathcal{K}\) is encoded by the same model as \(\mathbf{v}_{k_i}\).} The semantic similarity between a test case and the update is measured via cosine similarity:
  \[
  \text{sim}(t_j, \mathcal{K}) = \max_{k_i \in \mathcal{K}} \cos\left(\mathbf{v}_{t_j}, \mathbf{v}_{k_i}\right).
  \]

  While semantic similarity captures direct relevance, many bugs are caused by the interaction of new features with existing complex logic. To reflect the intrinsic test value of each case, we define a Semantic Complexity Score based on the diversity of interactions. Specifically, we normalize the counts of distinct actions, objects, scenes, UI components, and states into the range \([0,1]\), and compute their average. To prevent long paths from inflating the score, we further divide by the normalized path length:
  \[
  \text{SCS}(t_j) = \frac{\hat{A}_j + \hat{O}_j + \hat{S}_j + \hat{U}_j + \hat{G}_j}{\hat{L}_j},
  \]
  where \(\hat{A}_j, \hat{O}_j, \hat{S}_j, \hat{U}_j, \hat{G}_j\) denote the normalized counts of actions, objects, scenes, UI components, and states, respectively, and \(\hat{L}_j\) denotes the normalized path length.

  The final prioritization score combines both semantic relevance and complexity:
  \[
  \text{Score}(t_j) = \lambda \cdot \text{sim}(t_j, \mathcal{K}) + (1-\lambda) \cdot \frac{\text{SCS}(t_j)}{\max_{t \in \mathcal{T}} \text{SCS}(t)},
  \]
  where \(\lambda \in [0, 1]\) balances update relevance and functional complexity. This hybrid strategy ensures that high-priority cases not only cover newly changed content but also retain a potential to uncover critical bugs.

  \subsection{Test Case Execution and Maintenance}
  To ensure efficient regression testing, we adopt a lightweight incremental maintenance strategy. Each test case is re-executed on the latest game version and classified based on outcome:
  If the original goal is still achievable, the test case is marked as  \emph{valid} and its associated metadata in the test repository is updated accordingly.
  Otherwise, it is marked as \emph{obsolete} and flagged for manual review or regeneration.

  \section{Evaluation}
  \label{sec:evaluation}

  We aim to evaluate our proposed framework through three research questions:  
  \begin{itemize}
      \item \textbf{RQ1 (Effectiveness):} How well does \methodname{} detect and diversify bugs compared to baseline methods?
      \item \textbf{RQ2 (Efficiency):} How efficient is \methodname{} in performing regression testing during version updates?
      \item \textcolor{black}{\textbf{RQ3 (Ablation Analysis):} What are the respective contributions of the LLM-guided seed generation, RL-guided exploration, multi-objective optimization, and update-aware prioritization components?}
  \end{itemize}

  \subsection{Experimental Setup}

  \subsubsection{Game Environments}
  \label{sec:env_versions}

  \begin{figure}[ht]
  \centering
  \includegraphics[width=0.95\linewidth]{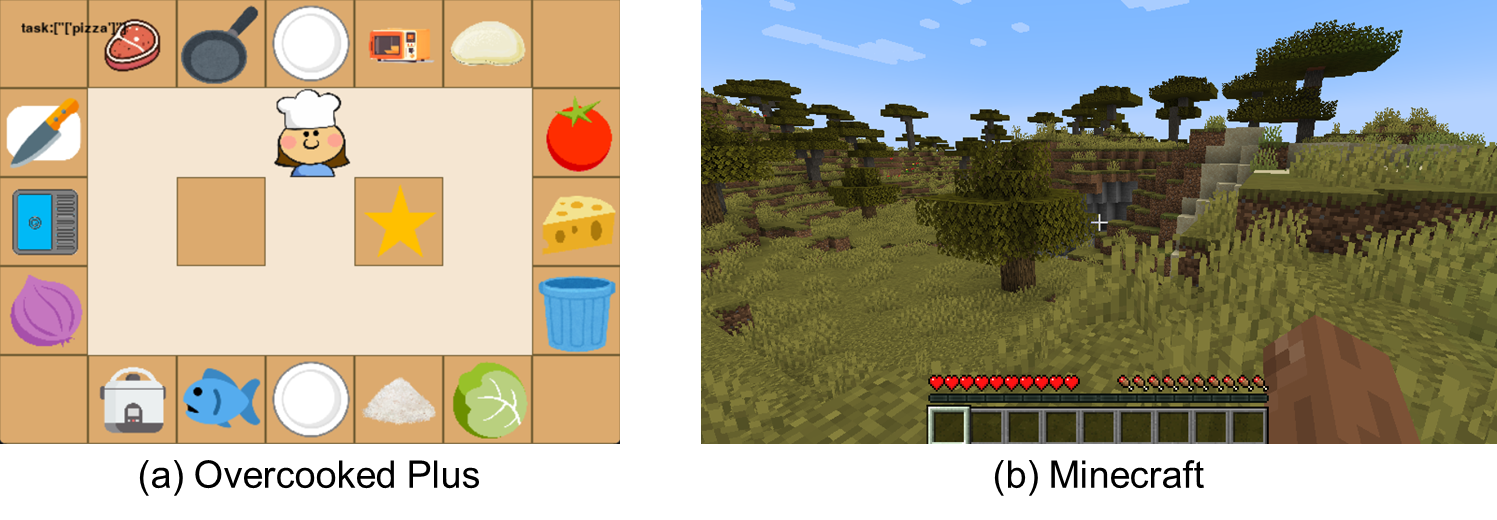}
  \caption{Illustration of game environments used in evaluation.}
  \label{fig:gameimage}
  \end{figure}

  We conduct experiments in two game environments: Overcooked Plus and Minecraft. 
  \textit{Overcooked Plus.}  
  Overcooked Plus~\cite{Cai2024OvercookedPlus} is an open-source reimplementation of the commercial game \textit{Overcooked}~\cite{overcooked_team17}. It preserves the original’s high task complexity while enhancing operational constraints and system controllability. This environment is particularly suitable for evaluating the execution accuracy and planning capability of testing methods in rule-driven games.  
  In this environment, players must complete sequential operations such as chopping, cooking, plating, and serving. The rules are strict and deterministic; for example, each workstation can only hold one item, and mistakes require explicit correction. The environment is fully observable, with the state and position of all objects represented in structured form, which facilitates behavioral modeling and analysis.  
  We designed 37 tasks and 2 version updates (V1$\rightarrow$V2 and V2$\rightarrow$V3). Specifically, V1 contains 10 tasks, and V2 extends to 27 tasks. Both updates not only introduce new game tasks but also include bug fixes and modifications to interaction logic. 
  During development, we conducted systematic and unbiased bug collection, ensuring that the recorded issues—spanning logic errors, interaction conflicts, UI or rendering issues, and occasional physics anomalies—reflect a representative distribution of real-world problems encountered during gameplay. Reproducible triggers were constructed for each bug to support rigorous regression testing. \textcolor{black}{Each recorded defect was assigned a unique bug ID and translated, before evaluation, into a bug oracle implemented as a reproducible trigger predicate. We validated each oracle through manual replay and cross-check by multiple authors against the intended game rules and update specification, to ensure that the trigger corresponded to a true defect rather than intended behavior.}
  This environment emphasizes precise sequential planning and error tolerance, making it well-suited for assessing the breadth of behavioral coverage and robustness of testing methods in rule-driven settings.

  \textit{Minecraft.}  
  Minecraft~\cite{minecraft2009} is a highly open-ended sandbox game with extremely rich interaction mechanisms and a vast state space, and has been widely adopted in AI research. This environment is suitable for evaluating the generalization ability and adaptive strategies of testing methods in open-world and exploratory tasks.  
  The Minecraft environment used in our study is based on a custom mod currently under development. The tasks within this mod are inspired by the original achievement system, covering multiple interaction types such as resource collection, crafting, and combat. To ensure compatibility across baselines, complex operations are abstracted into parameterized commands (e.g., “mine,” “craft,” “move”), which are then mapped to atomic low-level actions. The environment is also fully observable, exposing key variables such as character position, inventory state, available resources, and tool usage.  
  In this environment, we define 50 tasks and two version updates (V1$\rightarrow$V2 and V2$\rightarrow$V3), covering action space extensions, item accessibility adjustments, and quest logic modifications. All bugs recorded during development are retained and made reproducible through trigger mechanisms to ensure the authenticity and comparability of results. \textcolor{black}{The benchmark bugs mainly involve quest logic and trigger errors, resource or inventory inconsistencies, collision or pathfinding failures, and performance anomalies. As in Overcooked Plus, each retained defect is assigned a unique bug ID and a validated trigger-based oracle before evaluation; oracle validation is performed through manual replay and cross-check by multiple authors against the intended mod logic and update specification.}  
  This environment emphasizes long-term planning and strategic flexibility, making it suitable for evaluating the exploration breadth, behavioral diversity, and version adaptability of testing methods in open-world settings.

  \subsubsection{Evaluation Metrics}
  \label{sec:metrics}
  For effectiveness (RQ1), we consider \textit{Bug Count}, \textit{Unique Bugs}, \textit{Unique States}, \textit{Reward}, and \textit{Success Rate}.  
  \textit{Bug Count} measures the total number of bug activations under a fixed testing budget. \textcolor{black}{A bug occurrence is defined as an executed trajectory satisfying a pre-defined and validated bug oracle (e.g., illegal state transition, resource overflow, or inconsistent game logic).} Bug Count is reported as a global cumulative value across all episodes.  
  \textit{Unique Bugs} records the number of distinct bug IDs triggered at least once. \textcolor{black}{Each bug oracle is associated with a unique bug ID defined before evaluation, and each ID is counted only once even if its oracle is repeatedly triggered across multiple episodes.}  
  \textit{Unique States} represents the number of distinct environment states visited during testing, reflecting the behavioral diversity of generated test cases. \textcolor{black}{In the actual implementation, environment states are encoded as structured state representations (in Overcooked Plus: all object positions and player states; in Minecraft: player position, inventory contents, and local block configuration) and then globally deduplicated across all executed episodes.}  
  \textit{Reward} represents the cumulative in-game reward obtained per episode during each test case, while \textit{Success Rate} denotes the proportion of test cases that are marked as successful task completions under the environment's completion criterion.

  \textcolor{black}{Formally, let \(M\) denote the total number of executed episodes (test cases), \(T_e\) the number of steps in episode \(e\), \(s_{e,t}\) the visited state at step \(t\), \(r_{e,t}\) the immediate reward, and \(y_e \in \{0,1\}\) the episode-level success indicator. Then:}

  \[
  \textcolor{black}{V_e = \{s_{e,t}\mid t=1,\dots,T_e\}, \qquad
  \mathrm{Unique\ States} = \left|\bigcup_{e=1}^{M} V_e\right|.}
  \]

  \[
  \textcolor{black}{R_e = \sum_{t=1}^{T_e} r_{e,t}, \qquad
  \mathrm{Reward} = \frac{1}{M}\sum_{e=1}^{M} R_e.}
  \]

\[
\textcolor{black}{\mathrm{Success\ Rate} = \frac{1}{M}\sum_{e=1}^{M} y_e.}
\]

  For efficiency (RQ2), we record both \textit{Duration}, the total wall-clock time required to execute all test cases of a method\footnote{All experiments were conducted on a workstation equipped with an AMD Ryzen 9 9950X3D CPU, 64 GB RAM, and an NVIDIA RTX 5090 GPU running Ubuntu 22.04 under Windows 11 WSL.}, and \textit{Steps}, the cumulative number of action steps performed during testing.

  \subsubsection{Comparison Methods}
  \label{sec:baselines}
  We compare against five representative methods, described below:

  \textit{(i) Random~\cite{hu2024language,li2022gbgallery}.}
  A general-purpose exploration baseline based on uniform random action sampling over all available actions, with each candidate action assigned equal probability. Invalid actions are not excluded. This serves as a lower bound for uninformed exploration.

  \textit{(ii) PPO~\cite{PPO}.}
  As a state-of-the-art general-purpose exploration agent, we adopt the Stable-Baselines3~\cite{raffin2021stable} implementation. \textcolor{black}{Specifically, we use PPO with \texttt{MlpPolicy}. Unless otherwise stated, the policy follows the default SB3 MLP architecture for vector observations, using a \texttt{FlattenExtractor} and separate actor/critic networks with two hidden layers of 64 units each, \texttt{Tanh} activations, orthogonal initialization, and Adam optimization.} Models are pre-trained for 100{,}000 steps with learning rate $3\times 10^{-4}$, discount factor $\gamma=0.95$, and batch size 256. \textcolor{black}{The remaining PPO hyperparameters follow the SB3 defaults.} The resulting models are then stored in \texttt{data/models}. During inference, we load the corresponding model and execute it in stochastic mode (\texttt{deterministic = False}) for the prescribed number of steps, consistent with other baselines.
  \textcolor{black}{For fair comparison, all PPO-based components in our framework use the same Stable-Baselines3 implementation and the same PPO configuration unless otherwise stated.}

  \textit{(iii) diff-Qlearning~\cite{RLReg}.}
  In contrast to the general-purpose exploration methods above, diff-Qlearning is a regression-specific RL approach proposed by~\cite{RLReg}, specifically designed for detecting potential regressions in game environments. The method trains a Q-learning agent across different software versions, leveraging \textit{differential behavior signals} as implicit supervision to guide exploration toward areas likely to contain regressions. Unlike policy-based methods such as PPO, diff-Qlearning treats training trajectories themselves as test cases. Each episode is executed under both the old and new versions of the game, and behavioral discrepancies—such as differences in rewards, state transitions, or violations of pre-defined rules—are used to reinforce states that may indicate bugs introduced during version updates. Following the configuration described in the original paper, we implement diff-Qlearning using a learning rate $\alpha = 0.2$, discount factor $\gamma = 0.99$, and an initial exploration rate $\epsilon = 0.5$, which decays every 100 steps until reaching a minimum of 0.01. The total training step budget is kept consistent with other baselines to ensure fair comparison.

  \textit{(iv) Human-recorded test cases~\cite{ostrowski2013automated}.}
  To align with practical settings, we also include human-recorded test cases as high-quality human baselines~\cite{politowski2021survey}. Three experienced testers participated, each with 2–3 years of professional experience in game QA or development and over 200 hours of gameplay experience in the original versions of \textit{Overcooked} and \textit{Minecraft}. Each tester recorded five trajectories per task using keyboard and mouse, resulting in approximately 1,800 trajectories in total. On average, each version update required about 6 hours of recording per participant.

  \textit{(v) \methodname{} (ours).}
  \textcolor{black}{In the current implementation, the game runtime state is maintained in structured form, and task-relevant variables are extracted from it as the structured state representation used for node construction in the state--action graph. Before each LLM planning call, this structured state representation is converted through rule-based state-to-text templates into a concise natural-language observation, which is then combined with the task objective and available actions. The LLM outputs a sequence of executable actions for task completion.}
  For each environment, we prompt LLM to generate 20 seed trajectories per task. \textcolor{black}{These LLM-generated seeds constitute the initial training dataset for Behavior Cloning (BC),} and are used to train a behavior cloning policy, which initializes a PPO agent for guided exploration (using the same PPO configuration as above but with the task-specific reward design introduced below). The resulting exploration trajectories are aggregated into the candidate path pool, which is then optimized using the multi-objective procedure over cost, coverage, and rarity, with rarity measured via 2-gram subsequences. \textcolor{black}{During candidate-path extraction, we bound both path length and path enumeration for each task-completion terminal state. In the reported experiments, the extractor enumerates at most 1{,}000 shortest simple paths from the initial state to each terminal state, and keeps only those whose length does not exceed 50.}
  \textcolor{black}{For update-aware selection, the same LLM is then prompted to extract semantic tags from each update log before similarity scoring. We instantiate the Sentence-BERT encoder used for semantic similarity with the \texttt{all-MiniLM-L6-v2} checkpoint~\cite{sentence-transformers-all-minilm-l6-v2}, which is built on a MiniLM backbone~\cite{wang-etal-2020-minilm}. In all reported experiments, we use a default upper bound of 20 extracted semantic tags per update log. This setting keeps the change representation concise while limiting redundancy from overly fine-grained tags; if an update log contains fewer salient changed components, fewer tags are returned.}
  During experiments, we record different degrees of RTS proportion (from 10\% to 90\%), and unless otherwise specified, we use \methodname{} (top 50\%) as the default configuration for discussion to ensure consistency and clarity.  
  \textcolor{black}{\textit{Ablation.} In addition to the full version, we evaluate five ablated variants to analyze the contribution of each component along the pipeline. For the first generation stage, \emph{without LLM-guided seed generation} removes the LLM-generated seeds and the BC warm start, so PPO starts from random initialization while the downstream update-aware selection stage remains unchanged; \emph{without RL exploration} keeps the LLM-generated seeds but removes the subsequent PPO-guided exploration, constructing candidate tests only from those seed trajectories. For the downstream stages, \emph{without multi-objective optimization} removes the multi-objective optimization module, \emph{without RTS} removes the update-aware prioritization module, and \emph{without both} disables both downstream modules while retaining all generated outputs. Unless a removed component makes a specific step unavailable, all ablation settings follow the same environments, task sets, step budgets, PPO hyperparameters, and evaluation protocol as the full method.}
  We consistently used GPT-4o as the LLM model throughout the entire experimental process. \textcolor{black}{Prompt design was stage-specific. For seed generation, the prompt provides the current game situation, task objective, and available actions to elicit high-level executable plans; for update-aware selection, the prompt constrains the model to return concise gameplay-relevant semantic tags in a stable format. The final prompts were determined through pilot runs in both environments, with minor adjustments to the task instructions, context information, and output schema to improve executability and output stability.}

  To ensure fair and consistent evaluation, all learning-based methods (PPO, diff-Qlearning, and \methodname{}) share the same action and observation space, random seed (\texttt{seed=42}). For all reinforcement-learning-based methods, each episode and task follow a fixed step budget to control exploration cost. In Overcooked Plus, each episode is capped at 100 steps and each task at 5000 steps; in Minecraft, 75 steps per episode and 3000 per task. In Overcooked Plus, the reward design assigns +1000 for each stage goal \textcolor{black}{(e.g., completing required ingredient processing)}, +10000 for completing the overall quest, and -0.1 per action step. In Minecraft, the design assigns -0.5 per action step, +10 for intermediate goals (e.g., obtaining relevant items), and +100 for completing the quest. Each method is executed independently 10 times (except Human-recorded tests), and results are reported as the mean with 95\% confidence intervals. A two-tailed t-test is applied to assess statistical significance.

  \subsection{Experimental Results}
  %\jialong{这里图上粗体是什么意思？一般是性能最好的标粗，但这样的话就变成SAGE top 90\%粗了，所以我建议把粗体删了}
  % ==================== Overcooked Plus ====================

  \begin{table*}[ht]
  \centering
  \caption{Comparison of methods on V1$\rightarrow$V2 and V2$\rightarrow$V3 in Overcooked Plus (mean with standard deviation). \methodname{} conducts experiments on different RTS proportion and compiles the results.}
  \label{tab:overcooked}
  \resizebox{\textwidth}{!}{
  \begin{tabular}{llrrrrrrrr}
  \Xhline{1.2pt}
  \multirow{2.5}{*}{Version} & \multirow{2.5}{*}{Method} & \multicolumn{6}{c}{RQ1} & \multicolumn{2}{c}{RQ2} \\
  \cmidrule(lr){3-8} \cmidrule(lr){9-10}
  &  & Episodes & Bug Count & Unique Bugs & Unique States & Reward & Success Rate & Duration (s) & Total Steps \\
  \Xhline{0.8pt}

  \multirow{22}{*}{V1$\rightarrow$V2}
  & \multirow{2}{*}{RANDOM}                 & \makecell[r]{500.4 \\ ($\pm$0.5)} & \makecell[r]{5313.8 \\ ($\pm$33.8)} & \makecell[r]{23.9 \\ ($\pm$2.0)} & \makecell[r]{6920.1 \\ ($\pm$161.1)} & \makecell[r]{-491.2 \\ ($\pm$12.1)} & \makecell[r]{0.00 \\ ($\pm$0.00)} & \makecell[r]{14.5 \\ ($\pm$0.2)} & \makecell[r]{50000.0 \\ ($\pm$0.0)} \\
  \cmidrule(lr){2-10}
  & \multirow{2}{*}{PPO}                    & \makecell[r]{565.0 \\ ($\pm$37.1)} & \makecell[r]{25835.7 \\ ($\pm$4078.3)} & \makecell[r]{19.3 \\ ($\pm$3.1)} & \makecell[r]{218.6 \\ ($\pm$50.1)} & \makecell[r]{171.8 \\ ($\pm$86.9)} & \makecell[r]{0.18 \\ ($\pm$0.08)} & \makecell[r]{69.3 \\ ($\pm$1.3)} & \makecell[r]{50000.0 \\ ($\pm$0.0)} \\
  \cmidrule(lr){2-10}
  & \multirow{2}{*}{diff-Qlearning}         & \makecell[r]{523.2 \\ ($\pm$21.4)} & \makecell[r]{4722.2 \\ ($\pm$694.1)} & \makecell[r]{28.4 \\ ($\pm$3.1)} & \makecell[r]{2607.4 \\ ($\pm$208.3)} & \makecell[r]{-42.3 \\ ($\pm$13.2)} & \makecell[r]{0.00 \\ ($\pm$0.00)} & \makecell[r]{35.1 \\ ($\pm$0.5)} & \makecell[r]{50000.0 \\ ($\pm$0.0)} \\
  \cmidrule(lr){2-10}
  & HUMAN                  & 80.0 & 743.7 & 40.3 & 1490.0 & 1162.9 & 0.99 & 14.9 & 3038.7 \\
  \cmidrule(lr){2-10}
  & \makecell[l]{\methodname{} \\ (top 10\%)} & \makecell[r]{119.3 \\ ($\pm$8.0)} & \makecell[r]{974.4 \\ ($\pm$84.9)} & \makecell[r]{33.8 \\ ($\pm$2.1)} & \makecell[r]{1832.5 \\ ($\pm$128.7)} & \makecell[r]{1138.8 \\ ($\pm$22.4)} & \makecell[r]{1.00 \\ ($\pm$0.00)} & \makecell[r]{1.3 \\ ($\pm$0.1)} & \makecell[r]{4419.1 \\ ($\pm$289.1)} \\
  \cmidrule(lr){2-10}
  & \makecell[l]{\methodname{} \\ (top 30\%)} & \makecell[r]{357.0 \\ ($\pm$24.4)} & \makecell[r]{3000.9 \\ ($\pm$244.0)} & \makecell[r]{36.9 \\ ($\pm$2.4)} & \makecell[r]{2992.5 \\ ($\pm$178.6)} & \makecell[r]{1153.3 \\ ($\pm$14.8)} & \makecell[r]{1.00 \\ ($\pm$0.00)} & \makecell[r]{3.8 \\ ($\pm$0.3)} & \makecell[r]{13004.9 \\ ($\pm$893.0)} \\
  \cmidrule(lr){2-10}
  & \makecell[l]{\methodname{} \\ (top 50\%)} & \makecell[r]{594.3 \\ ($\pm$40.5)} & \makecell[r]{5038.5 \\ ($\pm$415.9)} & \makecell[r]{37.5 \\ ($\pm$2.3)} & \makecell[r]{3526.6 \\ ($\pm$200.2)} & \makecell[r]{1160.3 \\ ($\pm$11.6)} & \makecell[r]{1.00 \\ ($\pm$0.00)} & \makecell[r]{6.2 \\ ($\pm$0.5)} & \makecell[r]{21189.3 \\ ($\pm$1474.4)} \\
  \cmidrule(lr){2-10}
  & \makecell[l]{\methodname{} \\ (top 70\%)} & \makecell[r]{832.1 \\ ($\pm$56.7)} & \makecell[r]{7102.7 \\ ($\pm$581.5)} & \makecell[r]{38.2 \\ ($\pm$2.2)} & \makecell[r]{3769.6 \\ ($\pm$213.5)} & \makecell[r]{1163.0 \\ ($\pm$10.9)} & \makecell[r]{1.00 \\ ($\pm$0.00)} & \makecell[r]{8.5 \\ ($\pm$0.6)} & \makecell[r]{28913.3 \\ ($\pm$2073.8)} \\
  \cmidrule(lr){2-10}
  & \makecell[l]{\methodname{} \\ (top 90\%)} & \makecell[r]{1069.8 \\ ($\pm$73.1)} & \makecell[r]{9283.6 \\ ($\pm$749.6)} & \makecell[r]{38.3 \\ ($\pm$2.2)} & \makecell[r]{3840.3 \\ ($\pm$213.2)} & \makecell[r]{1167.8 \\ ($\pm$10.0)} & \makecell[r]{1.00 \\ ($\pm$0.00)} & \makecell[r]{10.8 \\ ($\pm$0.9)} & \makecell[r]{36276.6 \\ ($\pm$2700.7)} \\
  \Xhline{1.2pt}

  \multirow{22}{*}{V2$\rightarrow$V3}
  & \multirow{2}{*}{RANDOM}                 & \makecell[r]{1850.6 \\ ($\pm$0.7)} & \makecell[r]{10554.5 \\ ($\pm$64.5)} & \makecell[r]{35.4 \\ ($\pm$1.3)} & \makecell[r]{21531.3 \\ ($\pm$186.6)} & \makecell[r]{-415.0 \\ ($\pm$5.8)} & \makecell[r]{0.00 \\ ($\pm$0.00)} & \makecell[r]{58.1 \\ ($\pm$0.4)} & \makecell[r]{185000.0 \\ ($\pm$0.0)} \\
  \cmidrule(lr){2-10}
  & \multirow{2}{*}{PPO}                    & \makecell[r]{1975.0 \\ ($\pm$57.6)} & \makecell[r]{62712.9 \\ ($\pm$10990.9)} & \makecell[r]{26.4 \\ ($\pm$2.7)} & \makecell[r]{537.6 \\ ($\pm$67.6)} & \makecell[r]{76.6 \\ ($\pm$43.6)} & \makecell[r]{0.09 \\ ($\pm$0.04)} & \makecell[r]{253.2 \\ ($\pm$3.7)} & \makecell[r]{185000.0 \\ ($\pm$0.0)} \\
  \cmidrule(lr){2-10}
  & \multirow{2}{*}{diff-Qlearning}         & \makecell[r]{1955.1 \\ ($\pm$73.3)} & \makecell[r]{11263.8 \\ ($\pm$1264.0)} & \makecell[r]{38.3 \\ ($\pm$2.4)} & \makecell[r]{7484.2 \\ ($\pm$478.2)} & \makecell[r]{-27.0 \\ ($\pm$8.8)} & \makecell[r]{0.00 \\ ($\pm$0.00)} & \makecell[r]{135.9 \\ ($\pm$0.6)} & \makecell[r]{185000.0 \\ ($\pm$0.0)} \\
  \cmidrule(lr){2-10}
  & HUMAN                  & 224.7 & 2204.3 & 65.7 & 4972.0 & 1144.0 & 0.99 & 22.8 & 9134.3 \\
  \cmidrule(lr){2-10}
  & \makecell[l]{\methodname{} \\ (top 10\%)} & \makecell[r]{388.5 \\ ($\pm$16.0)} & \makecell[r]{3072.4 \\ ($\pm$117.4)} & \makecell[r]{53.7 \\ ($\pm$1.6)} & \makecell[r]{5449.1 \\ ($\pm$160.5)} & \makecell[r]{1167.5 \\ ($\pm$13.3)} & \makecell[r]{1.00 \\ ($\pm$0.00)} & \makecell[r]{3.3 \\ ($\pm$0.1)} & \makecell[r]{14404.7 \\ ($\pm$584.4)} \\
  \cmidrule(lr){2-10}
  & \makecell[l]{\methodname{} \\ (top 30\%)} & \makecell[r]{1164.9 \\ ($\pm$47.8)} & \makecell[r]{9220.0 \\ ($\pm$337.2)} & \makecell[r]{56.8 \\ ($\pm$1.7)} & \makecell[r]{8494.0 \\ ($\pm$245.4)} & \makecell[r]{1168.3 \\ ($\pm$12.3)} & \makecell[r]{1.00 \\ ($\pm$0.00)} & \makecell[r]{9.8 \\ ($\pm$0.5)} & \makecell[r]{42432.3 \\ ($\pm$1755.3)} \\
  \cmidrule(lr){2-10}
  & \makecell[l]{\methodname{} \\ (top 50\%)} & \makecell[r]{1941.0 \\ ($\pm$79.5)} & \makecell[r]{15173.9 \\ ($\pm$617.9)} & \makecell[r]{57.7 \\ ($\pm$2.0)} & \makecell[r]{9874.1 \\ ($\pm$254.9)} & \makecell[r]{1164.1 \\ ($\pm$12.2)} & \makecell[r]{1.00 \\ ($\pm$0.00)} & \makecell[r]{16.1 \\ ($\pm$0.9)} & \makecell[r]{69284.9 \\ ($\pm$2837.6)} \\
  \cmidrule(lr){2-10}
  & \makecell[l]{\methodname{} \\ (top 70\%)} & \makecell[r]{2717.5 \\ ($\pm$111.2)} & \makecell[r]{20975.7 \\ ($\pm$911.8)} & \makecell[r]{58.6 \\ ($\pm$2.0)} & \makecell[r]{10539.0 \\ ($\pm$269.0)} & \makecell[r]{1158.9 \\ ($\pm$13.4)} & \makecell[r]{1.00 \\ ($\pm$0.00)} & \makecell[r]{22.1 \\ ($\pm$1.2)} & \makecell[r]{95129.2 \\ ($\pm$4073.9)} \\
  \cmidrule(lr){2-10}
  & \makecell[l]{\methodname{} \\ (top 90\%)} & \makecell[r]{3493.9 \\ ($\pm$143.1)} & \makecell[r]{26602.6 \\ ($\pm$1055.3)} & \makecell[r]{58.6 \\ ($\pm$2.0)} & \makecell[r]{10793.0 \\ ($\pm$279.5)} & \makecell[r]{1153.8 \\ ($\pm$15.2)} & \makecell[r]{1.00 \\ ($\pm$0.00)} & \makecell[r]{27.9 \\ ($\pm$1.5)} & \makecell[r]{119422.9 \\ ($\pm$5084.1)} \\
  \Xhline{1.2pt}
  \end{tabular}}
  \end{table*}

  % ==================== Minecraft ====================
  \begin{table*}[ht]
  \centering
  \caption{Comparison of methods on V1$\rightarrow$V2 and V2$\rightarrow$V3 in Minecraft (mean with standard deviation). Rewards are normalized by Episodes (Reward/Episode).}
  \label{tab:minecraft}
  \resizebox{\textwidth}{!}{
  \begin{tabular}{llrrrrrrrr}
  \Xhline{1.2pt}
  \multirow{2.5}{*}{Version} & \multirow{2.5}{*}{Method} & \multicolumn{6}{c}{RQ1} & \multicolumn{2}{c}{RQ2} \\
  \cmidrule(lr){3-8} \cmidrule(lr){9-10}
  &  & Episodes & Bug Count & Unique Bugs & Unique States & Reward & Success Rate & Duration (s) & Total Steps \\
  \Xhline{0.8pt}

  \multirow{24}{*}{V1$\rightarrow$V2}
  & \multirow{2}{*}{RANDOM} & \makecell[r]{1652.9 \\ ($\pm$8.1)} & \makecell[r]{21726.3 \\ ($\pm$203.6)} & \makecell[r]{22.0 \\ ($\pm$2.7)} & \makecell[r]{5249.3 \\ ($\pm$135.7)} & \makecell[r]{-16.8 \\ ($\pm$0.8)} & \makecell[r]{0.06 \\ ($\pm$0.01)} & \makecell[r]{61.7 \\ ($\pm$3.1)} & \makecell[r]{120000 \\ ($\pm$0)} \\
  \cmidrule(lr){2-10}
  & \multirow{2}{*}{PPO} & \makecell[r]{1654.3 \\ ($\pm$5.6)} & \makecell[r]{21721.9 \\ ($\pm$267.1)} & \makecell[r]{23.3 \\ ($\pm$2.4)} & \makecell[r]{5623.3 \\ ($\pm$82.2)} & \makecell[r]{-15.4 \\ ($\pm$0.5)} & \makecell[r]{0.06 \\ ($\pm$0.05)} & \makecell[r]{269.3 \\ ($\pm$12.8)} & \makecell[r]{120000 \\ ($\pm$0)} \\
  \cmidrule(lr){2-10}
  & \multirow{2}{*}{diff-Qlearning} & \makecell[r]{1658.4 \\ ($\pm$8.3)} & \makecell[r]{21957.0 \\ ($\pm$209.7)} & \makecell[r]{23.3 \\ ($\pm$2.2)} & \makecell[r]{5524.9 \\ ($\pm$54.5)} & \makecell[r]{-15.1 \\ ($\pm$0.8)} & \makecell[r]{0.07 \\ ($\pm$0.01)} & \makecell[r]{389.0 \\ ($\pm$19.4)} & \makecell[r]{120000 \\ ($\pm$0)} \\
  \cmidrule(lr){2-10}
  & HUMAN & 200 & 783 & 41.3 & 1789 & 142.1 & 1.00 & 42.7 & 3857 \\
  \cmidrule(lr){2-10}
  & \makecell[l]{\methodname{} \\ (top 10\%)} & \makecell[r]{200.8 \\ ($\pm$26.8)} & \makecell[r]{9838.4 \\ ($\pm$1887.6)} & \makecell[r]{38.2 \\ ($\pm$2.7)} & \makecell[r]{7313.0 \\ ($\pm$1549.0)} & \makecell[r]{104.8 \\ ($\pm$14.9)} & \makecell[r]{1.00 \\ ($\pm$0.00)} & \makecell[r]{53.8 \\ ($\pm$2.7)} & \makecell[r]{19799.2 \\ ($\pm$3073.4)} \\
  \cmidrule(lr){2-10}
  & \makecell[l]{\methodname{} \\ (top 30\%)} & \makecell[r]{664.6 \\ ($\pm$57.6)} & \makecell[r]{20414.2 \\ ($\pm$3474.6)} & \makecell[r]{38.8 \\ ($\pm$1.6)} & \makecell[r]{16402.8 \\ ($\pm$2781.7)} & \makecell[r]{116.2 \\ ($\pm$10.4)} & \makecell[r]{1.00 \\ ($\pm$0.00)} & \makecell[r]{94.2 \\ ($\pm$4.7)} & \makecell[r]{49426.0 \\ ($\pm$6441.9)} \\
  \cmidrule(lr){2-10}
  & \makecell[l]{\methodname{} \\ (top 50\%)} & \makecell[r]{1069.4 \\ ($\pm$172.7)} & \makecell[r]{25031.6 \\ ($\pm$6397.6)} & \makecell[r]{39.0 \\ ($\pm$2.0)} & \makecell[r]{20201.6 \\ ($\pm$5250.4)} & \makecell[r]{122.7 \\ ($\pm$21.7)} & \makecell[r]{1.00 \\ ($\pm$0.00)} & \makecell[r]{134.6 \\ ($\pm$6.7)} & \makecell[r]{66094.0 \\ ($\pm$12996.1)} \\
  \cmidrule(lr){2-10}
  & \makecell[l]{\methodname{} \\ (top 70\%)} & \makecell[r]{1519.0 \\ ($\pm$186.1)} & \makecell[r]{29428.0 \\ ($\pm$7190.8)} & \makecell[r]{39.1 \\ ($\pm$2.0)} & \makecell[r]{22185.4 \\ ($\pm$3425.0)} & \makecell[r]{126.9 \\ ($\pm$15.4)} & \makecell[r]{1.00 \\ ($\pm$0.00)} & \makecell[r]{161.5 \\ ($\pm$8.0)} & \makecell[r]{80558.2 \\ ($\pm$13535.1)} \\
  \cmidrule(lr){2-10}
  & \makecell[l]{\methodname{} \\ (top 90\%)} & \makecell[r]{1974.0 \\ ($\pm$259.7)} & \makecell[r]{32151.6 \\ ($\pm$6658.4)} & \makecell[r]{39.2 \\ ($\pm$2.0)} & \makecell[r]{23280.6 \\ ($\pm$4514.0)} & \makecell[r]{130.5 \\ ($\pm$19.8)} & \makecell[r]{1.00 \\ ($\pm$0.00)} & \makecell[r]{182.7 \\ ($\pm$9.1)} & \makecell[r]{91234.4 \\ ($\pm$14579.7)} \\
  \Xhline{1.2pt}

  \multirow{24}{*}{V2$\rightarrow$V3}
  & \multirow{2}{*}{RANDOM} & \makecell[r]{3338.6 \\ ($\pm$11.0)} & \makecell[r]{43455.3 \\ ($\pm$208.1)} & \makecell[r]{24.0 \\ ($\pm$1.8)} & \makecell[r]{10510.4 \\ ($\pm$144.6)} & \makecell[r]{-15.0 \\ ($\pm$0.6)} & \makecell[r]{0.07 \\ ($\pm$0.05)} & \makecell[r]{123.5 \\ ($\pm$6.3)} & \makecell[r]{240000 \\ ($\pm$0)} \\
  \cmidrule(lr){2-10}
  & \multirow{2}{*}{PPO} & \makecell[r]{3361.1 \\ ($\pm$12.7)} & \makecell[r]{43713.1 \\ ($\pm$578.7)} & \makecell[r]{25.7 \\ ($\pm$1.2)} & \makecell[r]{11229.3 \\ ($\pm$113.0)} & \makecell[r]{-12.9 \\ ($\pm$0.4)} & \makecell[r]{0.08 \\ ($\pm$0.06)} & \makecell[r]{538.7 \\ ($\pm$26.3)} & \makecell[r]{240000 \\ ($\pm$0)} \\
  \cmidrule(lr){2-10}
  & \multirow{2}{*}{diff-Qlearning} & \makecell[r]{3362.6 \\ ($\pm$15.1)} & \makecell[r]{44001.9 \\ ($\pm$335.5)} & \makecell[r]{25.3 \\ ($\pm$2.3)} & \makecell[r]{10941.6 \\ ($\pm$82.1)} & \makecell[r]{-13.0 \\ ($\pm$0.7)} & \makecell[r]{0.08 \\ ($\pm$0.05)} & \makecell[r]{777.9 \\ ($\pm$38.9)} & \makecell[r]{240000 \\ ($\pm$0)} \\
  \cmidrule(lr){2-10}
  & HUMAN & 400 & 1896 & 45.4 & 4523 & 149.3 & 1.00 & 85.3 & 9836 \\
  \cmidrule(lr){2-10}
  & \makecell[l]{\methodname{} \\ (top 10\%)} & \makecell[r]{489.8 \\ ($\pm$58.6)} & \makecell[r]{31827.2 \\ ($\pm$3974.6)} & \makecell[r]{41.8 \\ ($\pm$1.4)} & \makecell[r]{24387.6 \\ ($\pm$2700.7)} & \makecell[r]{111.7 \\ ($\pm$13.5)} & \makecell[r]{1.00 \\ ($\pm$0.00)} & \makecell[r]{53.8 \\ ($\pm$2.7)} & \makecell[r]{58063.4 \\ ($\pm$7126.4)} \\
  \cmidrule(lr){2-10}
  & \makecell[l]{\methodname{} \\ (top 30\%)} & \makecell[r]{1604.0 \\ ($\pm$95.0)} & \makecell[r]{65699.0 \\ ($\pm$6657.8)} & \makecell[r]{42.0 \\ ($\pm$1.2)} & \makecell[r]{56541.4 \\ ($\pm$3207.5)} & \makecell[r]{125.5 \\ ($\pm$7.3)} & \makecell[r]{1.00 \\ ($\pm$0.00)} & \makecell[r]{94.2 \\ ($\pm$4.7)} & \makecell[r]{143468.8 \\ ($\pm$9028.6)} \\
  \cmidrule(lr){2-10}
  & \makecell[l]{\methodname{} \\ (top 50\%)} & \makecell[r]{2562.8 \\ ($\pm$243.9)} & \makecell[r]{78379.0 \\ ($\pm$5634.8)} & \makecell[r]{42.0 \\ ($\pm$1.2)} & \makecell[r]{69782.4 \\ ($\pm$6312.6)} & \makecell[r]{133.3 \\ ($\pm$13.8)} & \makecell[r]{1.00 \\ ($\pm$0.00)} & \makecell[r]{134.6 \\ ($\pm$6.7)} & \makecell[r]{188592.0 \\ ($\pm$14826.1)} \\
  \cmidrule(lr){2-10}
  & \makecell[l]{\methodname{} \\ (top 70\%)} & \makecell[r]{3647.2 \\ ($\pm$289.0)} & \makecell[r]{87927.6 \\ ($\pm$7485.1)} & \makecell[r]{42.2 \\ ($\pm$1.4)} & \makecell[r]{74639.4 \\ ($\pm$1929.9)} & \makecell[r]{139.5 \\ ($\pm$10.6)} & \makecell[r]{1.00 \\ ($\pm$0.00)} & \makecell[r]{161.5 \\ ($\pm$8.0)} & \makecell[r]{223544.6 \\ ($\pm$20127.8)} \\
  \cmidrule(lr){2-10}
  & \makecell[l]{\methodname{} \\ (top 90\%)} & \makecell[r]{4735.8 \\ ($\pm$463.4)} & \makecell[r]{95531.6 \\ ($\pm$12404.4)} & \makecell[r]{42.2 \\ ($\pm$1.4)} & \makecell[r]{76924.8 \\ ($\pm$7286.2)} & \makecell[r]{143.7 \\ ($\pm$14.9)} & \makecell[r]{1.00 \\ ($\pm$0.00)} & \makecell[r]{182.7 \\ ($\pm$9.1)} & \makecell[r]{250517.0 \\ ($\pm$21743.4)} \\
  \Xhline{1.2pt}
  \end{tabular}}
  \end{table*}

  \subsection{RQ1: Effectiveness}
  \label{sec:rq1}

  The experimental results are summarized in Table~\ref{tab:overcooked} and Table~\ref{tab:minecraft}.  
  In the Overcooked Plus environment, \methodname{} detected an average of 37.5 and 57.7 unique bugs in the two regression phases, closely approaching human-recorded trajectories (40.3 and 65.7) and far exceeding all automated baselines. 
  For instance, PPO and diff-Qlearning identified only around 19.3 and 28.4 unique bugs in the V1$\rightarrow$V2 phase, and 26.4 and 38.3 in the V2$\rightarrow$V3 phase—roughly half of \methodname{}’s diversity.  
  Notably, even though the number of unique states explored by \methodname{} is lower than Random exploration, it consistently achieves higher unique-bug diversity and a perfect success rate (1.0).

  A similar pattern is observed in the Minecraft environment, which presents a significantly larger and more complex state space.  
  While PPO and diff-Qlearning triggered over 20K total bug activations, they revealed fewer than 26 unique bugs, indicating repetitive activation of shallow, low-impact issues.  
  In contrast, \methodname{} uncovered 39 and 42 unique bugs in the two regression phases—about 1.6× more than the automated baseline—while maintaining competitive state diversity (20K–70K).  
  These results show that the proposed framework achieves robust generalization and semantic adaptability even under open-world conditions.
  Through the collaboration of LLM and RL, \methodname{} can stably complete tasks while leveraging its state–action graph to infer additional feasible paths, effectively generating more diverse and semantically rich test cases from limited exploration. This structured exploration paradigm grants it a unique advantage in uncovering deep logic regressions.

  \textit{Representative bug examples.}  
  To illustrate the types of bugs that \methodname{} detected while baselines missed, we highlight three representative cases that demonstrate the framework's ability to expose complex, multi-step interaction defects:

  \textbf{(1) Multi-stage cooking anomaly (Overcooked Plus):} When a partially cooked ingredient is removed from the pot and then returned, the internal cooking timer fails to restore correctly. Even when the cumulative cooking time satisfies the requirement, the final dish is incorrectly judged as failed. Baseline methods typically struggle to complete full multi-step cooking workflows and thus cannot trigger such cross-stage state inconsistencies. In contrast, \methodname{} systematically explores diverse cooking sequences through LLM-guided seed trajectory generation combined with graph-based path optimization, enabling it to construct complex multi-stage scenarios where ingredients are repeatedly transferred between different cooking stations.

  \textbf{(2) Plate state reset anomaly (Overcooked Plus):} When a plate contains two or more ingredients requiring multi-stage processing, all ingredient states are incorrectly reset to the same value, leading to reward calculation failure when the task is submitted. Such states are typically reachable only after long action sequences, while baseline methods primarily rely on short-sighted random or heuristic exploration and rarely generate trajectories long enough to place multiple processed ingredients on the same plate. \methodname{}'s multi-objective optimization explicitly rewards path diversity and coverage, enabling it to discover these rare but critical state combinations.

  \textbf{(3) Near-full inventory crafting defect (Minecraft):} When the player's inventory contains item A (not at maximum stack limit) and attempts to craft an item that produces multiple output items (e.g., 1 log crafting into 4 planks), excess output items directly disappear rather than being dropped or blocking the crafting operation. Baseline methods often adopt greedy exploration strategies and do not systematically test boundary conditions related to inventory management. \methodname{}'s semantic-aware test case selection mechanism prioritizes scenarios related to inventory-related update logs, and its coverage-driven optimization further promotes exploration of boundary conditions such as near-full inventory states.

  These examples demonstrate that \methodname{}'s combination of LLM-guided exploration, graph-based multi-objective optimization, and semantic-aware selection enables systematic discovery of state-dependent bugs that require long action sequences or specific state combinations to trigger—precisely the types of defects that simple random or heuristic methods are unlikely to encounter.

  % \textit{Observed limitations.}  
  In some cases, we observed that \methodname{} failed to capture certain shallow bugs that were detected by diff-Qlearning or Random. Two main factors contribute to this behavior:  
  (1) During graph construction, although exploration rewards encourage broad coverage, the dominant task-oriented rewards still bias the agent toward goal completion. As a result, some shallow or off-task states are underexplored, leading to missed edge-case bugs (e.g., transient state-switching anomalies). 
  (2) During test case extraction from the graph, path search depth and breadth must be constrained for computational feasibility\textcolor{black}{, including limits on cycle revisits to avoid excessively repetitive trajectories}. In complex scenarios such as Minecraft V2$\rightarrow$V3, the task length and world openness lead to very large graph structures, forcing pruning that excludes certain peripheral states. Consequently, a portion of off-task test trajectories—while visited during exploration—are omitted from the final test set. This trade-off becomes more pronounced as task complexity increases.  
  Overall, these limitations reflect a balance between task-oriented depth and state-space breadth: \methodname{} prioritizes deep exploration and high-value bug detection at the cost of reduced coverage of non-critical states.

  %\textbf{Discussion and insights.}  
  Taken together, these results reveal a key insight: the “quantity” and “depth” of bug detection are independent yet equally important dimensions. While brute-force exploration may yield higher bug counts, truly effective regression testing depends on semantic diversity and coverage depth. Leveraging LLM-guided semantics and multi-objective optimization, \methodname{} strikes a balance between efficiency and quality—systematically exploring diverse behaviors while focusing on representative, high-value trajectories.  
  Furthermore, its consistently high task completion rate enables the system to traverse deep interaction chains and uncover latent logic regressions across modules and states.  
  From a practical standpoint, this quality-oriented testing paradigm better aligns with the fundamental goal of regression testing.
  The essence of regression testing lies in verifying whether a system can preserve the correctness and consistency of its core functionalities after updates or evolution. Therefore, rather than pursuing broad yet shallow bug activations, it is more crucial to conduct semantically deep explorations of critical functional chains to uncover complex logical or interactive regressions.
  By prioritizing depth over breadth, this strategy yields testing outcomes that more faithfully reflect the real quality risks of an evolving system. It avoids the redundant outputs of massive, repetitive, and low-impact issues commonly produced by traditional automation, thus achieving a more efficient and targeted regression verification process.

  \begin{tcolorbox}[colback=gray!5!white,colframe=gray!40!black,title=Summary of RQ1:]
  \methodname{} exhibits near human-level bug diversity and robust generalization across both environments.
  It detects approximately 1.6× more unique bugs than the automated baselines and performs only slightly below human experts.
  Unlike baselines that mainly uncover shallow or repetitive issues, the consistently high task success rate enables \methodname{} to reach late-stage interactions and expose complex cross-module regressions.
  Overall, this ``goal-oriented testing approach that balances depth and breadth'' aligns more closely with the essence of regression testing—prioritizing the verification of critical functional chains after system evolution—thereby providing a more accurate and efficient assessment of real quality risks than approaches that merely pursue wide but shallow coverage.
  \end{tcolorbox}

  \subsection{RQ2: Efficiency}
  \label{sec:rq2}

  As shown in Table~\ref{tab:overcooked} and Table~\ref{tab:minecraft}, \methodname{} consistently demonstrates significant efficiency advantages over all automated baselines across both version updates (V1$\rightarrow$V2 and V2$\rightarrow$V3), second only to Human Record.  
  In the Overcooked Plus environment, \methodname{} completes each regression cycle using roughly 21–69K steps, whereas PPO, diff-Qlearning, and Random exhaust their full 50K–185K step budgets.  
  In terms of duration, \methodname{} requires only about 6–16 seconds per cycle, compared with 14–69 seconds for PPO and diff-Qlearning.  
  Despite the much smaller interaction budget, \methodname{} still achieves higher unique-bug diversity and maintains a 100\% task success rate, indicating that its exploration trajectories are both efficient and representative.

  The trend is similar in the Minecraft environment, which features a larger state space, higher task complexity, and greater exploration cost.  
  \methodname{} completes the regression process within approximately 66–189K steps, while all RL-based baselines require the full 120–240K steps.  
  During the V1$\rightarrow$V2 transition, \methodname{} takes about 134 seconds on average, compared to 269 seconds for PPO and 389 seconds for diff-Qlearning.  
  During V2$\rightarrow$V3, it completes the process in roughly 134 seconds, while PPO and diff-Qlearning require 538 and 778 seconds, respectively.  
  Overall, \methodname{} reduces execution time by roughly 75–90\% compared with automated baselines while maintaining superior bug diversity and success rates, demonstrating strong scalability and efficiency in large-scale environments.

  The differences in duration among methods mainly arise from their respective planning paradigms.  
  Traditional RL methods such as PPO and diff-Qlearning adopt an online planning approach, where policies must be queried or updated continuously during testing.  
  For instance, diff-Qlearning performs a Q-table lookup for every action decision and computes differential rewards across two environments, significantly increasing duration.  
  Although PPO exhibits more stable convergence, its reliance on neural policy inference leads to higher decision latency per inference. In addition, the aforementioned duration does not include the training overhead. Especially in regression testing scenarios involving frequent version iterations, PPO and similar model-dependent inference methods often require retraining or fine-tuning as the system updates, introducing additional time and computational costs that further undermine their practicality.

  \textcolor{black}{To complement the execution-only measurements above, we additionally estimated the offline preparation cost of the learning-based methods using the training-step budgets described in Section~4.1.3. For the Overcooked setting, PPO models are trained for 100{,}000 steps per task, and diff-Qlearning follows the same total training-step budget; for a coarse comparison, we therefore treat diff-Qlearning as having the same order-of-magnitude training cost as PPO. A 10{,}000-step timing sample gave 9.01 seconds for baseline PPO, which implies about 3{,}333.7 seconds (\(\approx\)55.6 minutes) to prepare all 37 tasks once; under the same approximation, diff-Qlearning has the same total offline training cost. For \methodname{}, a 10{,}000-step sample required 9.29 seconds for PPO-guided exploration training, 161.71 seconds for graph/path extraction, and 6.73 seconds for Pareto optimization. Scaling only the PPO-training component to the 100{,}000-step budget yields about 261.34 seconds per task and 9{,}669.58 seconds (\(\approx\)161.2 minutes, 2.69 hours) to prepare all 37 tasks once. We emphasize that this estimate for \methodname{} is not pure policy training alone; it additionally includes graph construction, candidate-path extraction, and multi-objective optimization. Using the same empirical scaling factor for Minecraft (1.76981, rounded down after multiplication), the corresponding offline cost becomes about 17{,}113 seconds (\(\approx\)285.2 minutes, 4.75 hours) for \methodname{} and 5{,}900 seconds (\(\approx\)98.3 minutes, 1.64 hours) for PPO/diff-Qlearning.}

  Although \methodname{} achieves outstanding efficiency during the execution phase, it incurs a certain amount of preprocessing overhead during graph construction and multi-objective optimization.  
  \textcolor{black}{We also measured the associated LLM usage. Under the Overcooked setting (37 generation calls + 2 ranking calls), the total usage is approximately 101{,}700 \(\pm\) 18{,}500 input tokens and 3{,}410 \(\pm\) 1{,}480 output tokens. Under the Minecraft setting (50 generation calls + 2 ranking calls), the total usage is approximately 229{,}300 \(\pm\) 12{,}500 input tokens and 4{,}580 \(\pm\) 2{,}000 output tokens. These API costs are also incurred offline and therefore do not affect the execution-phase duration reported in Table~\ref{tab:overcooked} and Table~\ref{tab:minecraft}.}
  However, since the constructed state–action graph and optimized test cases can be reused across multiple versions, this offline cost can be effectively amortized.  
  Furthermore, while Human Record achieves the highest raw execution efficiency, the manual recording of gameplay trajectories typically requires several hours per session, limiting scalability in iterative testing pipelines.

  To study the impact of the RTS proportion, we examined RTS proportions of 10\%, 30\%, 50\%, 70\%, and 90\% to explore the trade-off between testing depth and efficiency.  
  Across both Overcooked and Minecraft, small proportions (10\%–30\%) yield most of the bug diversity while requiring only a fraction of the steps and time.  For example, in Overcooked V1$\rightarrow$V2, selecting the top 10\% of paths produces about 33.8 unique bugs in roughly 4K steps and 1.3 seconds, whereas increasing to 30\% achieves around 36.9 unique bugs at 13K steps and 3.8 seconds.  Expanding the subset to 70\% or 90\% increases the step count to 29–36K and duration to 8–11 seconds, yet the number of unique bugs saturates near 38.  
  A similar pattern is observed in the V2$\rightarrow$V3 transition and in Minecraft: smaller selections reduce steps and runtime dramatically with modest reductions in unique bugs, while larger selections yield only incremental coverage gains at much higher cost.  These results underline that moderate selections (10\%–50\%) can provide rapid validation or a balanced trade-off between efficiency and coverage, whereas very large proportions offer minimal additional benefit and significant overhead.

  \begin{tcolorbox}[colback=gray!5!white,colframe=gray!40!black,title=Summary of RQ2:]
  \methodname{} achieves comparable or higher bug coverage while using only about 10–40\% of the interaction steps required by baseline methods and reduces average duration by roughly 75–90\%, ranking second only to Human Record.  
  Its efficiency advantage primarily derives from the synergy between graph-based multi-objective optimization and update-aware test case prioritization, which together minimize redundant executions.  
  Although the preparation stage incurs additional overhead, this cost can be amortized across multiple version updates, making \methodname{} highly efficient for iterative regression testing.  
  Moreover, varying the RTS proportion enables flexible trade-offs between efficiency and coverage, with smaller subsets (10–50\%) already achieving near-optimal balance under most conditions.
  \end{tcolorbox}

  \subsection{RQ3: Ablation Analysis}
  \label{sec:rq3}

\begin{figure}[ht]
      \centering
      \includegraphics[width=\linewidth]{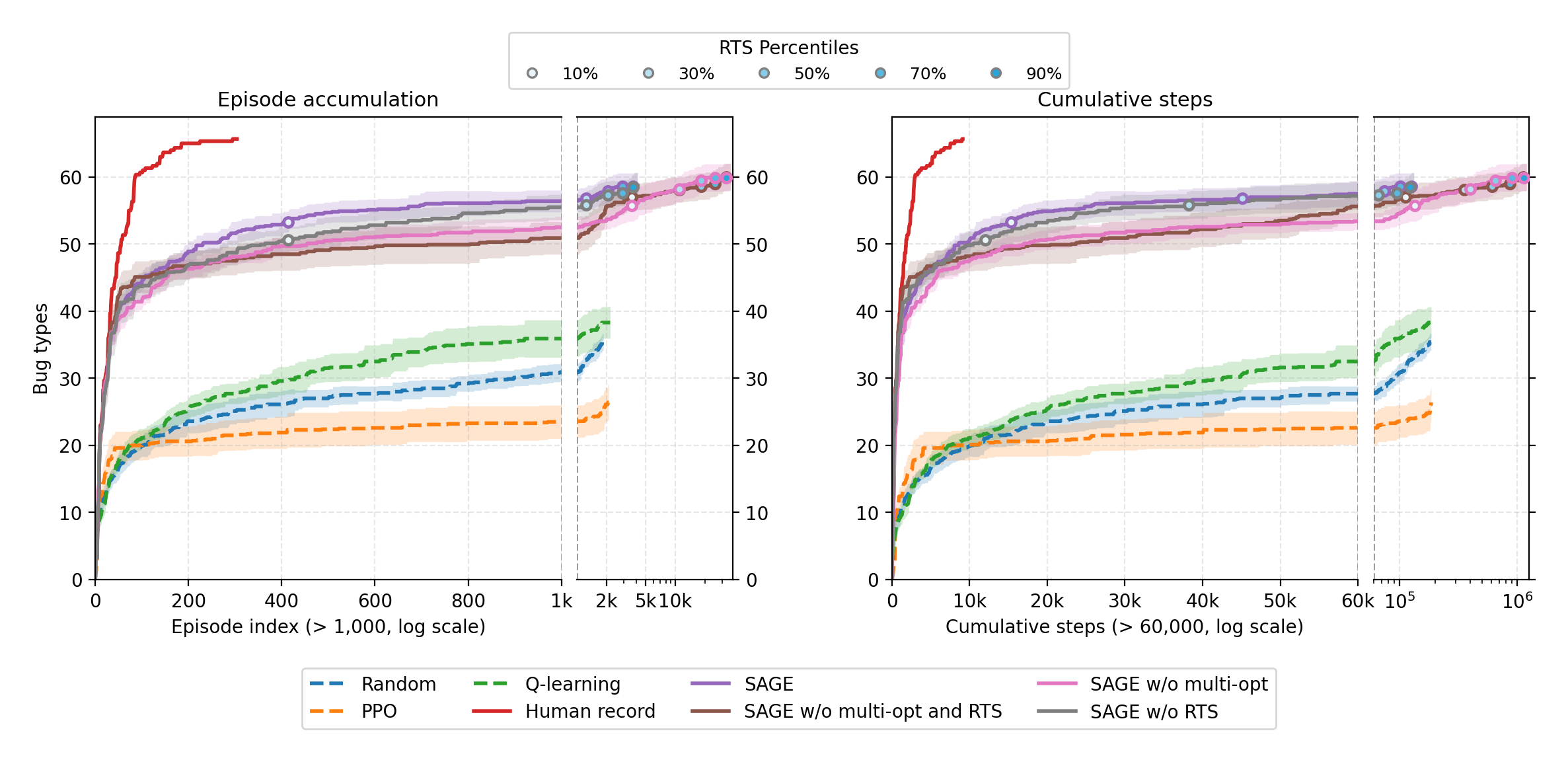}
      \caption{\textcolor{black}{Ablation study of the downstream optimization and prioritization modules of \methodname{} in Overcooked Plus. The figure shows the results of V2$\rightarrow$V3.}}
      \label{fig:ablation}
  \end{figure}

  {\color{black}To evaluate the contribution of the core components within \methodname{}, we organize the ablation analysis according to the pipeline order. We first examine the two subcomponents inside the generation stage---LLM-guided seed generation and RL-guided exploration---and then analyze the downstream multi-objective optimization and update-aware prioritization modules. Table~\ref{tab:foundation_ablation} reports the top-50\% results for the first-stage ablations in both Overcooked Plus and Minecraft, while Fig.~\ref{fig:ablation} visualizes the downstream-module ablations in Overcooked Plus V2$\rightarrow$V3.}

  {\color{black}
  \begin{table*}[ht]
  \centering
  \caption{Top-50\% ablation results for the generation-stage subcomponents of \methodname{} (mean with standard deviation).}
  \label{tab:foundation_ablation}
  \resizebox{\textwidth}{!}{
  \begin{tabular}{lllrrrrr}
  \Xhline{1.2pt}
  Environment & Version & Method & Episodes & Unique Bugs & Unique States & Success Rate & Total Steps \\
  \Xhline{0.8pt}
  \multirow{12}{*}{Overcooked Plus} & \multirow{6}{*}{V1$\rightarrow$V2} & \makecell[l]{\methodname{}} & \makecell[r]{594.3 \\ ($\pm$40.5)} & \makecell[r]{37.5 \\ ($\pm$2.3)} & \makecell[r]{3526.6 \\ ($\pm$200.2)} & \makecell[r]{1.00 \\ ($\pm$0.00)} & \makecell[r]{21189.3 \\ ($\pm$1474.4)} \\
  &  & w/o RL & \makecell[r]{189.8 \\ ($\pm$22.1)} & \makecell[r]{32.4 \\ ($\pm$2.0)} & \makecell[r]{825.0 \\ ($\pm$167.9)} & \makecell[r]{1.00 \\ ($\pm$0.00)} & \makecell[r]{6471.0 \\ ($\pm$1170.7)} \\
  &  & w/o LLM & \makecell[r]{437.0 \\ ($\pm$32.0)} & \makecell[r]{15.0 \\ ($\pm$1.0)} & \makecell[r]{1666.0 \\ ($\pm$110.0)} & \makecell[r]{1.00 \\ ($\pm$0.00)} & \makecell[r]{16493.0 \\ ($\pm$1180.0)} \\
  \cmidrule(lr){2-8}
  & \multirow{6}{*}{V2$\rightarrow$V3} & \makecell[l]{\methodname{}} & \makecell[r]{1941.0 \\ ($\pm$79.5)} & \makecell[r]{57.7 \\ ($\pm$2.0)} & \makecell[r]{9874.1 \\ ($\pm$254.9)} & \makecell[r]{1.00 \\ ($\pm$0.00)} & \makecell[r]{69284.9 \\ ($\pm$2837.6)} \\
  &  & w/o RL & \makecell[r]{269.7 \\ ($\pm$34.5)} & \makecell[r]{50.3 \\ ($\pm$2.1)} & \makecell[r]{1640.9 \\ ($\pm$206.8)} & \makecell[r]{1.00 \\ ($\pm$0.00)} & \makecell[r]{11857.2 \\ ($\pm$1490.7)} \\
  &  & w/o LLM & \makecell[r]{1023.0 \\ ($\pm$52.0)} & \makecell[r]{22.0 \\ ($\pm$1.0)} & \makecell[r]{3246.0 \\ ($\pm$120.0)} & \makecell[r]{1.00 \\ ($\pm$0.00)} & \makecell[r]{35931.0 \\ ($\pm$1700.0)} \\
  \cmidrule(lr){1-8}
  \multirow{12}{*}{Minecraft} & \multirow{6}{*}{V1$\rightarrow$V2} & \makecell[l]{\methodname{}} & \makecell[r]{1069.4 \\ ($\pm$172.7)} & \makecell[r]{39.0 \\ ($\pm$2.0)} & \makecell[r]{20201.6 \\ ($\pm$5250.4)} & \makecell[r]{1.00 \\ ($\pm$0.00)} & \makecell[r]{66094.0 \\ ($\pm$12996.1)} \\
  &  & w/o RL & \makecell[r]{386.6 \\ ($\pm$44.0)} & \makecell[r]{34.3 \\ ($\pm$3.3)} & \makecell[r]{4523.2 \\ ($\pm$1010.8)} & \makecell[r]{1.00 \\ ($\pm$0.00)} & \makecell[r]{21717.4 \\ ($\pm$2322.2)} \\
  &  & w/o LLM & \makecell[r]{776.9 \\ ($\pm$67.6)} & \makecell[r]{27.7 \\ ($\pm$2.6)} & \makecell[r]{11698.0 \\ ($\pm$1022.7)} & \makecell[r]{1.00 \\ ($\pm$0.00)} & \makecell[r]{42489.4 \\ ($\pm$6111.9)} \\
  \cmidrule(lr){2-8}
  & \multirow{6}{*}{V2$\rightarrow$V3} & \makecell[l]{\methodname{}} & \makecell[r]{2562.8 \\ ($\pm$243.9)} & \makecell[r]{42.0 \\ ($\pm$1.2)} & \makecell[r]{69782.4 \\ ($\pm$6312.6)} & \makecell[r]{1.00 \\ ($\pm$0.00)} & \makecell[r]{188592.0 \\ ($\pm$14826.1)} \\
  &  & w/o RL & \makecell[r]{421.0 \\ ($\pm$63.8)} & \makecell[r]{39.0 \\ ($\pm$1.5)} & \makecell[r]{10656.8 \\ ($\pm$1670.5)} & \makecell[r]{1.00 \\ ($\pm$0.00)} & \makecell[r]{36895.5 \\ ($\pm$3218.9)} \\
  &  & w/o LLM & \makecell[r]{2106.3 \\ ($\pm$259.7)} & \makecell[r]{30.1 \\ ($\pm$2.4)} & \makecell[r]{23100.0 \\ ($\pm$3917.9)} & \makecell[r]{1.00 \\ ($\pm$0.00)} & \makecell[r]{97567.9 \\ ($\pm$16121.8)} \\
  \Xhline{1.2pt}
  \end{tabular}}
  \end{table*}}

  {\color{black}The updated first-stage ablations reveal a stable ordering across both environments and both version updates. In terms of \emph{Unique Bugs}, the full method consistently performs best, followed by the variant without RL exploration and then the variant without LLM-guided seed generation. In terms of \emph{Episodes}, the full method also remains highest, while the variant without LLM-guided seed generation is consistently below it. This pattern indicates a clear division of labor between the two components: LLM-guided seed generation primarily improves the \emph{task feasibility} and semantic quality of trajectories, making more tasks successfully produce executable graphs and high-value regression tests, whereas RL exploration mainly expands the \emph{scale and breadth} of the explored state space once such trajectories are available.}

  {\color{black}A particularly important observation is that removing LLM guidance also reduces the number of executable test cases that survive into the final evaluation. This decrease in episodes is not mainly a replay-stage issue. Instead, a likely explanation from the observed pipeline behavior is that, in the pure-RL setting, fewer tasks successfully yield effective goal-reaching graphs from which valid paths can be extracted. As a result, the final RTS stage has fewer test cases available for execution. This effect is especially visible in Overcooked Plus, where the variant without LLM-guided seed generation drops from 594.3 to 437.0 episodes in V1$\rightarrow$V2 and from 1941.0 to 1023.0 in V2$\rightarrow$V3.}

  {\color{black}This contrast is especially pronounced in Overcooked Plus. Under the default top-50\% setting, removing LLM guidance reduces unique-bug coverage from 37.5 to 15.0 in V1$\rightarrow$V2 and from 57.7 to 22.0 in V2$\rightarrow$V3, while unique-state coverage remains at 1666.0 and 3246.0. By comparison, the variant without RL exploration retains much higher bug diversity---32.4 and 50.3 unique bugs---despite using far fewer episodes and covering a much smaller state space. This result shows that in a rule-intensive environment such as Overcooked Plus, simply exploring more states is not sufficient; semantically informed initialization is crucial for both constructing executable test cases and turning exploration into high-value regression tests.}

  {\color{black}Minecraft exhibits the same overall ordering, but with a milder degradation when LLM guidance is removed. In V1$\rightarrow$V2, the variant without LLM-guided seed generation still covers 11698.0 unique states across 776.9 episodes, yet detects only 27.7 unique bugs, compared with 34.3 for the variant without RL exploration and 39.0 for the full method. In V2$\rightarrow$V3, the same pattern persists: the variant without LLM-guided seed generation reaches 23100.0 unique states but remains at 30.1 unique bugs, whereas the variant without RL exploration achieves 39.0 unique bugs from only 10656.8 states. This again suggests that broad exploration alone does not guarantee effective regression testing; LLM-generated seeds improve whether exploration can be converted into valid, semantically meaningful test cases, while RL is responsible for amplifying that guidance into larger behavioral coverage.}

  {\color{black}The step statistics further clarify this trade-off. The variant without RL exploration is always the cheapest among the three first-stage settings, requiring the fewest episodes and interaction steps, while the full method is the most expensive because it continues exploring beyond the initial seed set. However, the extra cost consistently translates into the best combined bug diversity and state coverage. The fact that all extracted paths achieve a success rate of 1.00 also clarifies that the major difference among variants lies not in replay reliability after path extraction, but in whether the upstream generation stage can construct enough valid, goal-reaching trajectories in the first place.}

  {\color{black}After confirming the distinct roles of seed generation and RL exploration, we next analyze the downstream modules. Figure~\ref{fig:ablation} shows that the complete \methodname{} maintains the best balance between early bug discovery and overall coverage. Removing multi-objective optimization causes the coverage curve to saturate earlier, which indicates that many executed paths become behaviorally redundant and repeatedly trigger the same bugs. Removing update-aware prioritization produces a different effect: the final coverage remains competitive, but the system requires more episodes before reaching the same level of bug diversity. In other words, optimization mainly improves \emph{coverage efficiency} by compressing the candidate pool, whereas prioritization mainly improves \emph{ordering efficiency} by ensuring that the most relevant tests are executed first.}

  {\color{black}Taken together, the full ablation results reveal a coherent pipeline structure. LLM-guided seeds provide semantically meaningful starting behaviors, RL expands them into a broader behavioral space, multi-objective optimization removes redundant paths, and update-aware prioritization front-loads the most valuable test cases. The occasional observation that removing optimization can yield slightly higher final unique-bug coverage only after many more episodes further reinforces this interpretation: the optimization stage trades a small amount of long-tail coverage for a substantial gain in efficiency, which is a favorable compromise for practical regression testing under constrained budgets.}

  \begin{tcolorbox}[colback=gray!5!white,colframe=gray!40!black,title=Summary of RQ3:]
  \iffalse
  Both the multi-objective optimization and update-aware prioritization modules contribute critically to \methodname{}’s superior performance. 
  Optimization eliminates redundant trajectories and enhances the diversity and efficiency of generated test cases, while prioritization accelerates early bug discovery by executing high-value test cases first. 
  Together, these two modules balance efficiency and coverage, enabling effective and scalable regression testing under limited resources.
  \fi
  \textcolor{black}{The ablation study reveals a clear division of labor inside \methodname{}. LLM-guided seed generation is critical for producing semantically meaningful and executable task-oriented trajectories; removing it not only reduces bug diversity but also decreases the number of test cases that can be carried into the final RTS stage. RL exploration mainly serves as a coverage amplifier, substantially expanding the scale of explored behaviors and state-space breadth once effective seeds are available. Downstream, multi-objective optimization reduces redundancy among candidate paths, while update-aware prioritization improves early bug-discovery efficiency. Together, these components form a coherent pipeline that balances test quality, coverage breadth, and regression-testing efficiency.}
  \end{tcolorbox}

  \section{Discussion and Limitations}
  Beyond \methodname{}, the value of human-recorded test cases is also noteworthy. Although their coverage range is limited, they consistently exhibited high success rates and well-structured paths, reflecting the intuition and efficiency of experienced testers. This explains why human testing remains indispensable in practice and also motivates us to examine where automation still falls short.  

  Although our method has achieved promising results, several limitations remain that warrant further discussion.  
  First, the effectiveness of RL in complex and high-dimensional game environments remains a fundamental challenge. In some practical scenarios, RL agents may fail to complete tasks or to explore sufficiently diverse state spaces, thereby reducing the quality of the constructed state–action graphs. While prior studies have shown that LLMs demonstrate strong reasoning abilities in complex task decomposition, fully replacing RL with LLM-generated action sequences or relying on LLM self-correction mechanisms remains an open question. As a supplementary solution, incorporating human-recorded trajectories to assist graph construction is feasible in practice. Given the modular design of our framework, such alternatives can be flexibly integrated, but their effectiveness in practical environments still requires further validation.  

  Second, the multi-objective optimization stage depends on the availability of sufficient configuration and logging information from the game environment. In research prototypes, such information is often directly accessible. However, in more restricted grey-box environments, where internal system details are only partially observable, detailed debugging metrics are rarely available. A potential alternative is to capture screen recordings (e.g., via OBS) or use lightweight scripts to observe gameplay and infer environment changes through frame differencing, thereby augmenting the statistical features. Nevertheless, this approach introduces additional engineering overhead, and its accuracy and scalability remain open challenges.

  Third, our framework may incur non-trivial computational costs. In particular, the path extraction step from the constructed graph can become infeasible when the graph grows excessively large in complex games, leading to state explosion and resource exhaustion. This issue may hinder deployment in real-world contexts. The current mitigations include imposing limits on path search depth and maximum path counts, while future work could incorporate further pruning strategies to preserve behavioral diversity while reducing computational overhead.
  Moreover, although LLMs provide strong reasoning ability, their inference itself also introduces noticeable computational overhead and latency, which should be considered when deploying the framework at scale.

  \textcolor{black}{Fourth, beyond size considerations, three key technological constraints shape the applicability of SAGE. \textit{Observability constraint.} SAGE relies on game environments that expose structured runtime information through APIs, callbacks, or event logs. For fully black-box or strongly obfuscated scenarios, state extraction becomes unreliable, which limits graph fidelity. \textit{Abstraction granularity constraint.} The abstraction level must remain task-relevant: overly coarse abstraction merges behaviorally distinct states or actions, while overly fine-grained abstraction introduces sparsity and noise. Therefore, different game genres typically require tailored abstraction designs; the current semantic granularity may be insufficient for games with simple interactions or heavy real-time physics reliance (e.g., FPS, racing games). \textit{Action-space constraint.} For environments with continuous control or high-frequency physics interactions, actions must be discretized or parameterized before being represented in the unified graph form. Without such preprocessing, transition relationships become difficult to encode consistently. At the same time, these assumptions are still reasonable in many commercial game-development settings, where QA pipelines typically have access to runtime interfaces or logs, executable action abstractions, and human-written update logs or patch notes. As a result, SAGE is best suited for games that (1) expose structured runtime information through APIs or logs, (2) exhibit rich, distinguishable interaction elements across multiple semantic dimensions, and (3) feature action spaces that can be meaningfully abstracted into discrete or parameterized operations.}

  Finally, over long-term iterations, the gradual obsolescence of test cases may result in test suite drift, thereby reducing the effectiveness of regression testing. Therefore, test suite regeneration and manual intervention are still required. Future work should further explore more efficient incremental maintenance mechanisms to minimize redundant computation and ensure the long-term scalability of the framework in high-frequency, live-service game environments.

  \section{Threats to Validity}
  \subsection{Internal validity}  
  First, the training of RL agents in \methodname{} is inherently stochastic, as it depends on random initialization and exploration trajectories. Different random seeds may lead to noticeable performance variations. To mitigate this threat, we execute each experiment with multiple seeds and report averaged results together with standard deviations or confidence intervals.  

  Second, the performance of baseline methods can be highly sensitive to the design of reward functions and hyperparameter configurations. Suboptimal tuning may result in unfair comparisons and underestimate their true potential. To minimize this risk, we designed competitive reward functions and parameter settings in line with best practices.

  Third, our evaluation relies on \textcolor{black}{pre-defined, validated bug oracles derived from defects recorded during environment development or preserved from actual encountered issues. This design yields a controlled and repeatable benchmark for comparing methods on their ability to rediscover known regressions. However, it does not address the separate problem of automatically generating or validating oracles for previously unseen bugs in open-ended real-world deployments, and the resulting oracle set may not fully cover the diversity of bugs encountered in practice. In our current benchmark, and in many practical industrial workflows, oracle construction and confirmation still rely heavily on manual specification or manual validation. At the same time, recent research has started to explore how this process can be further automated, including LLM-based test oracle automation~\cite{10.1145/3715107} and game-specific oracle design in open-world settings such as Minecraft~\cite{11262358}.} 

  \subsection{External validity}  
  A first concern lies in the representativeness of the evaluated environments. Although our experiments cover Overcooked Plus and Minecraft, which represent rule-driven and open-world game genres respectively, \textcolor{black}{games as a broad software category encompass an extremely diverse range of genres that differ fundamentally from one another—from real-time competitive games to turn-based strategy, from single-player narrative experiences to massively multiplayer online worlds. Each genre exhibits distinct technical implementations, interaction patterns, and testing requirements.} Our evaluation does not encompass other mainstream types such as first-person shooters (FPS), \textcolor{black}{multiplayer online battle arenas (MOBAs),} card games, narrative-driven adventures, or large-scale multiplayer titles. \textcolor{black}{For instance, FPS games heavily rely on real-time responsiveness and physics simulation, where the current state abstraction granularity may be insufficient; card games feature relatively deterministic interactions but require different state space abstractions based on game rules and randomness; MOBA games involve complex multi-agent coordination and high-dimensional state spaces that may challenge both RL training efficiency and semantic feature extraction.} As a result, it is difficult to guarantee that our approach will generalize equally well across all categories of real-world games. \textcolor{black}{Additionally, the Minecraft environment, as a custom mod under active development, may have different internal development practices and a resulting bug profile compared to a mature commercial title.}

  The update logs used in our experiments were designed with reference to publicly available logs on commercial gaming platforms such as Steam\cite{steam2025}, covering common update categories (new content, gameplay adjustments, bug fixes). While this setup provides representative and controlled inputs for evaluation, we acknowledge that real production logs can be much noisier and more heterogeneous, with inconsistent formatting, redundancy, mixed terminology, and uneven granularity. In practice, SAGE can partially tolerate such variation through the LLM-based semantic extraction stage, and for highly noisy cases an additional LLM-based preprocessing step can be introduced before tag extraction to parse and structure raw logs into a more regularized form, as suggested by prior studies on LLM-based log parsing \cite{llmLogPreprocess2024}. We discuss this assumption and mitigation strategy in Section 5, and we leave large-scale validation on highly noisy industrial logs for future work.

  Finally, our evaluation primarily measures bug detection by counts and diversity without distinguishing the severity or business impact of individual bugs. In practical contexts, however, the ability to prioritize critical bugs may be more important than maximizing the sheer number of detected bugs. This mismatch between academic metrics and practical priorities also poses a potential external validity threat.

  \section{Conclusion and Future Work}
  \label{sec:conclusion}
  This paper presents \methodname{}, a unified framework for automated gray-box regression testing in games. Driven by semantic information, the framework leverages LLMs to systematically address three core issues. It solves the Foundation Issue using LLM-guided RL exploration to build a diverse test suite. It addresses the Maintenance Issue via a semantic-based multi-objective optimization to refine the suite into a compact, high-value subset. Finally, it tackles the Selection Issue by using an LLM to interpret update logs and translate changes into dynamic test priorities.

  We conducted a comprehensive evaluation in two representative game environments, comparing \methodname{} with both automated baselines and human-recorded test cases. The results demonstrate that our approach achieves superior bug detection capability and test case diversity, while reducing execution cost by a large margin compared to baselines. \textcolor{black}{Moreover, our ablation studies confirm the distinct contributions of LLM-guided seed generation, RL-guided exploration, multi-objective optimization, and update-aware prioritization, showing how these components jointly enhance regression testing performance.}

  For future work, we plan to extend our study in the following directions.
  First, we plan to extend our evaluation to \textcolor{black}{a broader range of game genres and} practical \textcolor{black}{development} settings. \textcolor{black}{The current experiments were conducted in two representative but limited game environments. Extending SAGE to additional game types—particularly those with different state observability characteristics (e.g., partially observable multiplayer games) or different semantic structures (e.g., real-time competitive games such as FPS and MOBAs, turn-based strategy games, card games with stochastic elements)—represents a critical direction for validating the generalizability of our approach. Moreover, collaborating} \textcolor{black}{with industry partners will allow us to assess the practical utility of our framework in real production pipelines, where tasks and update logs may be noisier, more heterogeneous, and larger-scale than in controlled research settings.}
  Second, we aim to further reduce the computational overhead of our method. While the proposed multi-objective optimization significantly improves test case quality, the path extraction and evaluation process can still become expensive in large state–action graphs. A promising direction is to integrate pruning strategies into the optimization stage, so that redundant or low-value paths can be efficiently discarded without sacrificing behavioral diversity. Such hybrid optimization–pruning techniques could greatly enhance the scalability and responsiveness of our approach in resource-constrained real-world workflows.
  Third, we will explore graph-based incremental maintenance mechanisms to improve long-term adaptability across version updates. Our current strategy focuses on re-execution and removal of obsolete test cases, but this process can become costly over extended development cycles. By explicitly modeling version-to-version transitions as update-aware graphs, we aim to incrementally adapt test suites with minimal recomputation. This approach has the potential to maintain consistency and stability of regression testing pipelines while reducing manual intervention and long-term overhead.

  \section*{Acknowledgement}
  This work was conducted during the author's visit to Southwest University.
  This research was partially supported by JSPS KAKENHI (No. 23K28064, 25K15290) and JST SPRING(No. JPMJSP2128).  

  \section*{Data Accessibility}
  The complete replication suite for the Overcooked Plus environment is publicly available at our project repository: \url{https://github.com/BlueLinkX/SAGE}
  . This includes the environment's source code, task configurations, \textcolor{black}{validated bug-oracle definitions and trigger implementations}, and all generated metadata used in our evaluation.
  For Minecraft, the source code is not public as the experiments were conducted using a custom modification (mod) that relies on Mojang/Microsoft's proprietary APIs. However, the code can be shared upon reasonable request for academic use.

  \appendix
  \section{Appendix I: Example Update Log}
  \label{appendix:update_log}

  Appendix~\ref{update_example} presents an illustrative example of an update document, showcasing the general format of version descriptions used in our experiments.

  \begin{tcolorbox}[
      colback=gray!5!white,
      colframe=black!50,
      title=\textbf{Example Update Log: Overcooked Plus (Version 2.0)},
      fonttitle=\bfseries,
  ]
  \label{update_example}
  \textbf{Release Date:} September 1, 2025

  \textbf{New Features:}
  \begin{itemize}
      \item \textit{Cheese Cutting Mechanic:} Cheese must now be chopped before use, introducing a new \texttt{chopped\_cheese} state.
      \item \textit{Soup Crafting System:} Added two new recipes, \texttt{MeatSoup} and \texttt{FishSoup}, with multi-ingredient dependencies (e.g., tomato + onion + meat/fish).
      \item ...
  \end{itemize}

  \textbf{Feature Changes:}
  \begin{itemize}
      \item Unified inheritance between \texttt{Fish} and \texttt{Meat} classes for consistent cooking logic.
      \item Adjusted interaction priorities between plates and cooking devices to reduce ambiguity.
      \item ...
  \end{itemize}

  \textbf{New Tasks:}
  \begin{itemize}
      \item \textit{Cheese Pizza:} Involves baking dough, slicing cheese, and assembling with tomato sauce before serving.
      \item \textit{Combo Meal Challenge:} A multi-step mission combining soup, baked bread, and grilled meat under strict time constraints.
      \item ...
  \end{itemize}

  \textbf{Bug Fixes:}
  \begin{itemize}
      \item Fixed an issue where chopped onions occasionally disappeared when placed on cutting boards.
      \item Fixed visual glitch where burned pizza retained its uncooked texture.
      \item ...
      
  \textbf{...}
  \end{itemize}
  \end{tcolorbox}

  \section{Appendix B: Example LLM Prompts}
  \label{appendix:llm_prompts}

  This appendix provides representative examples of the prompts employed in \methodname{}, illustrating how seed trajectories are generated for RL training and how large language models are guided to extract semantic tags from update logs.

  \begin{lstlisting}[language=json,firstnumber=1,caption={Example JSON prompt for generating a seed trajectory}]
  {
    "environment": {
      "name": "Overcooked Plus",
      "game_description": "This is a cooperative cooking simulation game where players must coordinate to prepare meals under time constraints...",
      "basic_rules": "Each ingredient must be processed correctly before serving...",
      \textcolor{black}{"current_obs_description": "[Converted from structured state representation] Kitchen layout: dough at (1,3) not yet kneaded; tomatoes at (2,1) not yet chopped; cheese at (3,2) ready; oven at (4,4) idle; serving table at (0,0) empty; player currently at (2,3) holding tomato.",}
      "available_actions": [
        "move_up", "move_down", "move_left", "move_right",
        "pickup", "drop", "interact", ...
      ]
    },

    "task": {
      "name": "Make Pizza",
      "task_objective": "Combine processed dough, tomato, and cheese into a pizza and bake it to completion...",
      "related_rules": "Dough must be kneaded before use; tomatoes must be chopped; cheese must be sliced..."
    },

    "past_solutions": [
      {
        "summary": ...,
        "key_subtasks": ...
      }
    ],

    "instructions": "Generate a new plan different from past_solutions to complete the task. If no past_solutions are available, ignore that field. Decompose the task into subtasks (e.g., fetch ingredients, chop, assemble, bake, serve). For each step, state its goal and the atomic action from available_actions. Ensure logical consistency and strategy diversity (e.g., different order, shorter path, or parallel workflow).",

    "output_format": "Return a JSON object containing: 1. summary: overall description of the new plan; 2. key_steps: list of main subtasks; 3. steps: an array of {step, description, action}."
  }
  \end{lstlisting}

  {\color{black}
  \begin{lstlisting}[language=json,firstnumber=1,caption={Example JSON prompt for generating a seed trajectory in Minecraft}]
  {
    "environment": {
      "name": "Minecraft",
      "game_description": "An open-ended sandbox game where players gather resources, craft tools, build structures, and complete quests. State is fully observable with access to player position, inventory, surrounding blocks, and NPC interactions.",
      "basic_rules": "Resources are mined from blocks and collected in inventory. Items can be crafted from recipes using collected materials. Some items require specific tools to mine (e.g., iron pickaxe for ore). Quests provide objectives and reward progress.",
      "current_obs_description": "[Converted from structured state representation] Player position: (120, 64, 180) in grassland biome. Inventory: wooden pickaxe, 32 logs, 8 planks, 1 iron ore. Surrounding resources: oak trees nearby at (125, 64, 185), stone deposits at (110, 60, 175), unstarted quest 'Craft Iron Tools' from NPC at (100, 64, 200).",
      "available_actions": [
        "move(north)", "move(south)", "move(east)", "move(west)", "move(up)", "move(down)",
        "mine(block_type)", "craft(item_name)", "pickup(item)", "drop(item)", "interact(npc)", ...
      ]
    },

    "task": {
      "name": "Craft Iron Tools",
      "task_objective": "Obtain at least one iron pickaxe by mining iron ore, smelting it into iron ingots, and crafting tools from the ingots. Deliver the iron pickaxe to the NPC at the quest marker.",
      "related_rules": "Iron ore requires a stone pickaxe or better to mine. Raw iron ore must be smelted in a furnace to produce iron ingots. Crafting recipes: 3 planks form a crafting table; 3 iron ingots plus 2 sticks form an iron pickaxe."
    },

    "past_solutions": [
      {
        "summary": "Collect logs, craft planks and sticks, mine stone with wooden pickaxe, craft stone pickaxe, mine iron ore, smelt iron, craft iron pickaxe.",
        "key_subtasks": ["gather logs", "craft sticks", "make crafting table", "craft stone pickaxe", "mine iron ore", "smelt iron", "craft iron pickaxe"]
      }
    ],

    "instructions": "Generate a new sequence of actions different from past_solutions to complete the task. If no past_solutions are available, ignore that field. Consider alternative approaches (e.g., finding pre-existing structure with furnace, alternative mining paths). For each step, provide the action from available_actions and a brief description. Ensure actions are grounded in the game mechanics and current state.",

    "output_format": "Return a JSON object containing: 1. summary: overall description of the new strategy; 2. key_steps: list of main objectives; 3. steps: an array of {step, action, description}."
  }
  \end{lstlisting}
  }

  {\color{black}
  \begin{lstlisting}[language=json,firstnumber=1,caption={Example prompt for extracting semantic tags from an update log}]
  {
    "update log": "Release Date: September 1, 2025\n New Features:....",

    "instructions": "Read the given update log text and identify up to 20 semantic tags representing affected game components, including but not limited to: items, actions, UI, functions, environment, and mechanics. Keep only the most salient tags. Do not extract version numbers or filler words. Each tag should be a lowercase phrase.",

    "output_format": "Return a JSON object with the following field: tags, a list of extracted semantic tags. If a brief summary is also returned for readability, it is auxiliary only and is not used in similarity scoring."
  }
  \end{lstlisting}
  }

  %\jialong{参考文献整理一下，有些地方数据缺失了，有些地方显示了2遍doi，另外正式出版的文献把doi信息都给加上显得正式一点}

  \bibliography{bib}

@INPROCEEDINGS{RLReg,
  author={Wu, Yuechen and Chen, Yingfeng and Xie, Xiaofei and Yu, Bing and Fan, Changjie and Ma, Lei},
  booktitle={2020 IEEE International Conference on Software Maintenance and Evolution (ICSME)}, 
  title={Regression Testing of Massively Multiplayer Online Role-Playing Games}, 
  year={2020},
  volume={},
  number={},
  pages={692-696},
  doi={10.1109/ICSME46990.2020.00074}}

@article{ostrowski2013automated,
  author    = {Michail Ostrowski and Samir Aroudj},
  title     = {Automated Regression Testing within Video Game Development},
  journal   = {GSTF Journal on Computing (JoC)},
  year      = {2013},
  volume    = {3},
  number    = {2},
  pages     = {10},
  doi       = {10.7603/s40601-013-0010-4},
  url       = {https://doi.org/10.7603/s40601-013-0010-4},
  issn      = {2010-2283}
}

@INPROCEEDINGS{GameRTS,
  author={Yu, Jiongchi and Wu, Yuechen and Xie, Xiaofei and Le, Wei and Ma, Lei and Chen, Yingfeng and Hu, Jingyu and Zhang, Fan},
  booktitle={2023 IEEE/ACM 45th International Conference on Software Engineering (ICSE)}, 
  title={GameRTS: A Regression Testing Framework for Video Games}, 
  year={2023},
  volume={},
  number={},
  pages={1393-1404},
  doi={10.1109/ICSE48619.2023.00122}}

@article{Alagarsamy2023,
title = {A3Test: Assertion-Augmented Automated Test case generation},
journal = {Information and Software Technology},
volume = {176},
pages = {107565},
year = {2024},
issn = {0950-5849},
doi = {https://doi.org/10.1016/j.infsof.2024.107565},
url = {https://www.sciencedirect.com/science/article/pii/S0950584924001708},
author = {Saranya Alagarsamy and Chakkrit Tantithamthavorn and Aldeida Aleti},
}

@INPROCEEDINGS{mioto2025mapping,
  author={Mioto, Vinícius and Petrillo, Fabio},
  booktitle={2025 IEEE/ACM 9th International Workshop on Games and Software Engineering (GAS)}, 
  title={A Mapping of Recording-based Game Test Automation Tools}, 
  year={2025},
  volume={},
  number={},
  pages={1-8},
  doi={10.1109/GAS66647.2025.00006}}

@inproceedings{de2014hybrid,
  author    = {De Souza, Luciano S. and Prud{\^e}ncio, Ricardo B.~C. and Barros, Fl{\'a}vio d.~A.},
  title     = {A Hybrid Binary Multi-Objective Particle Swarm Optimization with Local Search for Test Case Selection},
  booktitle = {2014 Brazilian Conference on Intelligent Systems (BRACIS)},
  pages     = {414--419},
  year      = {2014},
  doi       = {10.1109/BRACIS.2014.80}
}

@inproceedings{mondal2015exploring,
  author    = {Mondal, Dipankar and Hemmati, Hadi and Durocher, St{\'e}phane},
  title     = {Exploring Test Suite Diversification and Code Coverage in Multi-Objective Test Case Selection},
  booktitle = {2015 IEEE 8th International Conference on Software Testing, Verification and Validation (ICST)},
  pages     = {1--10},
  year      = {2015},
  doi       = {10.1109/ICST.2015.7102588}
}

@inproceedings{de2011multi,
  author={Souza, Luciano S. de and Miranda, Pericles B. C. de and Prudencio, Ricardo B. C. and Barros, Flavia de A.},
  booktitle={2011 IEEE 23rd International Conference on Tools with Artificial Intelligence}, 
  title={A Multi-objective Particle Swarm Optimization for Test Case Selection Based on Functional Requirements Coverage and Execution Effort}, 
  year={2011},
  volume={},
  number={},
  pages={245-252},
  doi={10.1109/ICTAI.2011.45}}

@article{spronck2006adaptive,
  author    = {Pieter Spronck and Marc Ponsen and Ida Sprinkhuizen-Kuyper and Eric Postma},
  title     = {Adaptive game AI with dynamic scripting},
  journal   = {Machine Learning},
  year      = {2006},
  volume    = {63},
  number    = {3},
  pages     = {217--248},
  doi       = {10.1007/s10994-006-6205-6},
  url       = {https://doi.org/10.1007/s10994-006-6205-6},
  issn      = {1573-0565}
}

@inproceedings{zheng2019wuji,
  author={Zheng, Yan and Xie, Xiaofei and Su, Ting and Ma, Lei and Hao, Jianye and Meng, Zhaopeng and Liu, Yang and Shen, Ruimin and Chen, Yingfeng and Fan, Changjie},
  booktitle={2019 34th IEEE/ACM International Conference on Automated Software Engineering (ASE)}, 
  title={Wuji: Automatic Online Combat Game Testing Using Evolutionary Deep Reinforcement Learning}, 
  year={2019},
  volume={},
  number={},
  pages={772-784},
  doi={10.1109/ASE.2019.00077}}

@inproceedings{guerrero-romero_using_2018,
  author={Guerrero-Romero, Cristina and Lucas, Simon M. and Perez-Liebana, Diego},
  booktitle={2018 IEEE Conference on Computational Intelligence and Games (CIG)}, 
  title={Using a Team of General AI Algorithms to Assist Game Design and Testing}, 
  year={2018},
  volume={},
  number={},
  pages={1-8},
  doi={10.1109/CIG.2018.8490417}}

@article{DBLP:journals/sbcjis/DuarteMDNE24,
author = {Duarte, Yohan and Canella, Henrique and Durelli, Vinicius and Nardi, Paulo and Endo, Andre},
year = {2024},
month = {07},
pages = {657–669},
title = {Exploratory testing for platform video games: strategies and lessons learned},
volume = {15},
journal = {Journal on Interactive Systems},
doi = {10.5753/jis.2024.4156}
}

@article{DBLP:conf/dsa/ZhangZAI23,
author="Mingyue, Zhang and Xiao-Yi, Zhang and Paolo, Arcaini and Fuyuki, Ishikawa",
title="An Investigation of the Behaviours of Machine Learning Agents Used in the Game of Go",
journal="2023 10th International Conference on Dependable Systems and Their Applications (DSA)",
publisher="IEEE",
year="2023",
month="08",
pages="734-742",
DOI="10.1109/dsa59317.2023.00105",
URL="https://cir.nii.ac.jp/crid/1360584339778405504"
}

@misc{overcooked_team17,
  title = {Overcooked!},
  author = {{Ghost Town Games Ltd.}},
  organization = {Team17},
  year = {2024},
  url = {https://www.team17.com/games/overcooked/},
  note = {Accessed: 2024-08-26},
  urldate = {2024-08-26}
}

@article{li2022gbgallery,
  author  = {Li, Z. and Wu, Y. and Ma, L. and Xie, X. and Chen, Y. and Fan, C.},
  title   = {GBGallery: A Benchmark and Framework for Game Testing},
  journal = {Empirical Software Engineering},
  volume  = {27},
  number  = {6},
  pages   = {Article 140},
  year    = {2022},
  doi     = {10.1007/s10664-022-10158-x}
}

@misc{minecraft2009,
  title        = {{Minecraft}},
  author       = {{Mojang Studios}},
  year         = {2009},
  howpublished = {\url{https://www.minecraft.net}},
  note         = {Sandbox video game developed by Mojang Studios}
}

@inproceedings{Cai2024OvercookedPlus,
  author={Cai, Jinyu and Li, Jialong and Li, Nianyu and Zhang, Mingyue and Yang, Ruijia and Tei, Kenji},
  booktitle={2024 IEEE International Conference on Autonomic Computing and Self-Organizing Systems Companion (ACSOS-C)}, 
  title={Overcooked Plus: A Comprehensive Cooking Scenario TestBed for Enhancing the Evaluation of Autonomous Planning Algorithms}, 
  year={2024},
  volume={},
  number={},
  pages={146-151},
  doi={10.1109/ACSOS-C63493.2024.00046}}

@misc{newzoo2024,
  title = {Newzoo's Global Games Market Report 2024 - Free Version},
  author = {{Newzoo}},
  year = {2024},
  url = {https://newzoo.com/resources/trend-reports/newzoos-global-games-market-report-2024-free-version},
  note = {Accessed: 2024-10-29}
}

@article{raffin2021stable,
  title={Stable-baselines3: Reliable reinforcement learning implementations},
  author={Raffin, Antonin and Hill, Ashley and Gleave, Adam and Kanervisto, Anssi and Ernestus, Maximilian and Dormann, Noah},
  journal={Journal of machine learning research},
  volume={22},
  number={268},
  pages={1--8},
  year={2021}
}

@inproceedings{hu2024language,
  author={Hu, Jie and Zhang, Mingyue and Liu, Bo and Wu, Yuechen and Chen, Yingfeng},
  booktitle={2024 IEEE 35th International Symposium on Software Reliability Engineering Workshops (ISSREW)}, 
  title={A Language-guided Acceleration Method for Smoke Testing of Game Quests}, 
  year={2024},
  volume={},
  number={},
  pages={7-12},
  doi={10.1109/ISSREW63542.2024.00039}}

@misc{PPO,
      title={Proximal Policy Optimization Algorithms}, 
      author={John Schulman and Filip Wolski and Prafulla Dhariwal and Alec Radford and Oleg Klimov},
      year={2017},
      eprint={1707.06347},
      archivePrefix={arXiv},
      primaryClass={cs.LG},
      url={https://arxiv.org/abs/1707.06347}, 
}

@INPROCEEDINGS{iftikhar2015automated,
  author={Iftikhar, Sidra and Iqbal, Muhammad Zohaib and Khan, Muhammad Uzair and Mahmood, Wardah},
  booktitle={2015 ACM/IEEE 18th International Conference on Model Driven Engineering Languages and Systems (MODELS)}, 
  title={An automated model based testing approach for platform games}, 
  year={2015},
  volume={},
  number={},
  pages={426-435},
  doi={10.1109/MODELS.2015.7338274}}

@inproceedings{gligoric2015practical,
author = {Gligoric, Milos and Eloussi, Lamyaa and Marinov, Darko},
title = {Practical regression test selection with dynamic file dependencies},
year = {2015},
isbn = {9781450336208},
publisher = {Association for Computing Machinery},
address = {New York, NY, USA},
doi = {10.1145/2771783.2771784},
booktitle = {Proceedings of the 2015 International Symposium on Software Testing and Analysis},
pages = {211–222},
numpages = {12},
location = {Baltimore, MD, USA},
series = {ISSTA 2015}
}

@INPROCEEDINGS{agrawal1993incremental,
  author={Agrawal, H. and Horgan, J.R. and Krauser, E.W. and London, S.A.},
  booktitle={1993 Conference on Software Maintenance}, 
  title={Incremental regression testing}, 
  year={1993},
  volume={},
  number={},
  pages={348-357},
  doi={10.1109/ICSM.1993.366927}}

@article{politowski2021survey,
title={A Survey of Video Game Testing},
author={Cristiano Politowski and Fábio Petrillo and Yann-Gaël Guéhéneuc},
journal={IEEE/ACM International Conference on Automation of Software Test (AST)},
year={2021},
pages={90-99},
doi={10.1109/AST52587.2021.00018}
}

@INPROCEEDINGS{9619048,
  author={Gordillo, Camilo and Bergdahl, Joakim and Tollmar, Konrad and Gisslén, Linus},
  booktitle={2021 IEEE Conference on Games (CoG)}, 
  title={Improving Playtesting Coverage via Curiosity Driven Reinforcement Learning Agents}, 
  year={2021},
  volume={},
  number={},
  pages={1-8},
  doi={10.1109/CoG52621.2021.9619048}}

@inproceedings {netravali2015mahimahi,
author = {Ravi Netravali and Anirudh Sivaraman and Somak Das and Ameesh Goyal and Keith Winstein and James Mickens and Hari Balakrishnan},
title = {Mahimahi: Accurate {Record-and-Replay} for {HTTP}},
booktitle = {2015 USENIX Annual Technical Conference (USENIX ATC 15)},
year = {2015},
isbn = {978-1-931971-225},
address = {Santa Clara, CA},
pages = {417--429},
url = {https://www.usenix.org/conference/atc15/technical-session/presentation/netravali},
publisher = {USENIX Association},
month = jul
}

@inproceedings{stahlke2020artificial,
author = {Stahlke, Samantha and Nova, Atiya and Mirza-Babaei, Pejman},
title = {Artificial Players in the Design Process: Developing an Automated Testing Tool for Game Level and World Design},
year = {2020},
isbn = {9781450380744},
publisher = {Association for Computing Machinery},
address = {New York, NY, USA},
url = {https://doi.org/10.1145/3410404.3414249},
doi = {10.1145/3410404.3414249},
booktitle = {Proceedings of the Annual Symposium on Computer-Human Interaction in Play},
pages = {267–280},
numpages = {14},
location = {Virtual Event, Canada},
series = {CHI PLAY '20}
}

@inproceedings{stahlke2019artificial,
author = {Stahlke, Samantha . and Nova, Atiya and Mirza-Babaei, Pejman},
title = {Artificial Playfulness: A Tool for Automated Agent-Based Playtesting},
year = {2019},
isbn = {9781450359719},
publisher = {Association for Computing Machinery},
address = {New York, NY, USA},
doi = {10.1145/3290607.3313039},
booktitle = {Extended Abstracts of the 2019 CHI Conference on Human Factors in Computing Systems},
pages = {1–6},
numpages = {6},
location = {Glasgow, Scotland Uk},
series = {CHI EA '19}
}

@misc{ixie2024gametesting,
  author       = {{iXie Gaming}},
  title        = {A Comprehensive Review of Game Test Automation Tools},
  year         = {2024},
  url          = {https://www.ixiegaming.com/blog/comprehensive-review-game-test-automation-tools/},
  note         = {Accessed: 2025-10-14},
}

@article{wang2024software,
  author={Wang, Junjie and Huang, Yuchao and Chen, Chunyang and Liu, Zhe and Wang, Song and Wang, Qing},
  journal={IEEE Transactions on Software Engineering}, 
  title={Software Testing With Large Language Models: Survey, Landscape, and Vision}, 
  year={2024},
  volume={50},
  number={4},
  pages={911-936},
  doi={10.1109/TSE.2024.3368208}}

@article{Tufano2020,
  title={Unit Test Case Generation with Transformers and Focal Context},
  author={Michele Tufano and others},
  journal={arXiv preprint arXiv:2009.05617},
  year={2020}
}

@inproceedings{TufanoAsserts2022,
author = {Tufano, Michele and Drain, Dawn and Svyatkovskiy, Alexey and Sundaresan, Neel},
title = {Generating accurate assert statements for unit test cases using pretrained transformers},
year = {2022},
isbn = {9781450392860},
publisher = {Association for Computing Machinery},
address = {New York, NY, USA},
url = {https://doi.org/10.1145/3524481.3527220},
doi = {10.1145/3524481.3527220},
booktitle = {Proceedings of the 3rd ACM/IEEE International Conference on Automation of Software Test},
pages = {54–64},
numpages = {11},
location = {Pittsburgh, Pennsylvania},
series = {AST '22}
}

@article{Xie2023,
  title={ChatUniTest: A ChatGPT-based Automated Unit Test Generation Tool},
  author={Xie, Yizhuo and others},
  journal={arXiv preprint arXiv:2305.04764},
  year={2023}
}

@article{Dakhel2023,
title = {Effective test generation using pre-trained Large Language Models and mutation testing},
journal = {Information and Software Technology},
volume = {171},
pages = {107468},
year = {2024},
issn = {0950-5849},
doi = {https://doi.org/10.1016/j.infsof.2024.107468},
url = {https://www.sciencedirect.com/science/article/pii/S0950584924000739},
author = {Arghavan Moradi Dakhel and Amin Nikanjam and Vahid Majdinasab and Foutse Khomh and Michel C. Desmarais},
}

@article{Mastropaolo2023,
  author  = {Mastropaolo, Antonio and Cooper, Nikita and Palacio, Diego~N. and Scalabrino, Simone and Poshyvanyk, Denys and Oliveto, Rocco and Bavota, Gabriele},
  title   = {Using Transfer Learning for Code-Related Tasks},
  journal = {IEEE Transactions on Software Engineering},
  volume  = {49},
  number  = {4},
  pages   = {1580--1598},
  year    = {2022},
  doi     = {10.1109/TSE.2022.3183297}
}

@inproceedings{Nashid2023,
  author    = {Nashid, Noor and Sintaha, Mifta and Mesbah, Ali},
  title     = {Retrieval-Based Prompt Selection for Code-Related Few-Shot Learning},
  booktitle = {Proceedings of the 45th IEEE/ACM International Conference on Software Engineering (ICSE)},
  pages     = {2450--2462},
  year      = {2023},
  doi       = {10.1109/ICSE48619.2023.00205}
}

@inproceedings{LiuFill2022,
  author    = {Liu, Zhe and Chen, Chunyang and Wang, Junjie and Che, Xing and Huang, Yuekai and Hu, Jun and Wang, Qing},
  title     = {Fill in the Blank: Context-Aware Automated Text Input Generation for Mobile GUI Testing},
  booktitle = {2023 IEEE/ACM 45th International Conference on Software Engineering (ICSE)},
  pages     = {1355--1367},
  year      = {2023},
  doi       = {10.1109/ICSE48619.2023.00119}
}

@inproceedings{LiuTestingExpert2023,
author = {Liu, Zhe and Chen, Chunyang and Wang, Junjie and Chen, Mengzhuo and Wu, Boyu and Che, Xing and Wang, Dandan and Wang, Qing},
title = {Make LLM a Testing Expert: Bringing Human-like Interaction to Mobile GUI Testing via Functionality-aware Decisions},
year = {2024},
isbn = {9798400702174},
publisher = {Association for Computing Machinery},
address = {New York, NY, USA},
doi = {10.1145/3597503.3639180},
booktitle = {Proceedings of the IEEE/ACM 46th International Conference on Software Engineering},
articleno = {100},
numpages = {13},
location = {Lisbon, Portugal},
series = {ICSE '24}
}

@article{DengEdgeCase2023,
  title={Large Language Models Are Edge-Case Fuzzers: Testing Deep Learning Libraries via FuzzGPT},
  author={Deng, Zhi and others},
  journal={arXiv preprint arXiv:2304.02014},
  year={2023}
}

@inproceedings{DengZeroShot2022,
  author    = {Deng, Yuyang and Xia, C.~S. and Peng, Han and Yang, Chuan and Zhang, Li},
  title     = {Large Language Models Are Zero-Shot Fuzzers: Fuzzing Deep-Learning Libraries via Large Language Models},
  booktitle = {Proceedings of the 32nd ACM SIGSOFT International Symposium on Software Testing and Analysis (ISSTA)},
  pages     = {423--435},
  year      = {2023},
  doi       = {10.1145/3597926.3598067}
}

@inproceedings{ZhangITiger2022,
  author    = {Zhang, Tianyi and Irsan, I.~C. and Thung, Ferdian and Han, Ding and Lo, David and Jiang, Lingxiao},
  title     = {iTiger: An Automatic Issue Title Generation Tool},
  booktitle = {Proceedings of the 30th ACM Joint European Software Engineering Conference and Symposium on the Foundations of Software Engineering (ESEC/FSE)},
  pages     = {1637--1641},
  year      = {2022},
  doi       = {10.1145/3540250.3558934}
}

@article{BuiDLR2022,
  author    = {Bui, Nghi D.~Q. and Wang, Yue and Hoi, Steven C.~H.},
  title     = {Detect--Localize--Repair: A Unified Framework for Learning to Debug with CodeT5},
  booktitle = {Findings of the Association for Computational Linguistics: EMNLP 2022},
  pages     = {812--823},
  year      = {2022},
  doi       = {10.18653/v1/2022.findings-emnlp.57}
}

@article{ChenSelfDebug2023,
  title={Teaching Large Language Models to Self-Debug},
  author={Chen, Shiqi and others},
  journal={arXiv preprint arXiv:2304.05128},
  year={2023}
}

@inproceedings{FengAdbGPT2023,
  author    = {Feng, Shen and Chen, Chun},
  title     = {Prompting Is All You Need: Automated Android Bug Replay with Large Language Models},
  booktitle = {Proceedings of the 46th IEEE/ACM International Conference on Software Engineering (ICSE)},
  pages     = {1--13},
  year      = {2024},
  doi       = {10.1145/3597503.3608137}
}

@inproceedings{Lajko2022,
author = {Lajk\'{o}, M\'{a}rk and Csuvik, Viktor and Vid\'{a}cs, L\'{a}szl\'{o}},
title = {Towards JavaScript program repair with generative pre-trained transformer (GPT-2)},
year = {2022},
isbn = {9781450392853},
publisher = {Association for Computing Machinery},
address = {New York, NY, USA},
doi = {10.1145/3524459.3527350},
booktitle = {Proceedings of the Third International Workshop on Automated Program Repair},
pages = {61–68},
numpages = {8},
location = {Pittsburgh, Pennsylvania},
series = {APR '22}
}

@inproceedings{WangExamples2023,
  author    = {Wang, Weishi and Wang, Yue and Joty, Shafiq and Hoi, Steven C.~H.},
  title     = {RAP-Gen: Retrieval-Augmented Patch Generation with CodeT5 for Automatic Program Repair},
  booktitle = {Proceedings of the 31st ACM Joint European Software Engineering Conference and Symposium on the Foundations of Software Engineering (ESEC/FSE)},
  pages     = {146--158},
  year      = {2023},
  doi       = {10.1145/3611643.3616256}
}

@misc{steam2025,
  author       = {{Valve Corporation}},
  title        = {Steam — The Ultimate Online Game Platform},
  year         = {2025},
  url          = {https://store.steampowered.com/},
  note         = {Accessed on October 30, 2025}
}

@article{10.1145/3715107,
author = {Molina, Facundo and Gorla, Alessandra and d’Amorim, Marcelo},
title = {Test Oracle Automation in the Era of LLMs},
year = {2025},
issue_date = {June 2025},
publisher = {Association for Computing Machinery},
address = {New York, NY, USA},
volume = {34},
number = {5},
issn = {1049-331X},
url = {https://doi.org/10.1145/3715107},
doi = {10.1145/3715107},
journal = {ACM Trans. Softw. Eng. Methodol.},
month = may,
articleno = {150},
numpages = {24}
}

@INPROCEEDINGS{11262358,
  author={Chen, Quan and Zhang, Mingyue and Zhang, Xiao–Yi},
  booktitle={2025 IEEE 36th International Symposium on Software Reliability Engineering Workshops (ISSREW)}, 
  title={Testing Open-World Games: A Minecraft Case Study}, 
  year={2025},
  volume={},
  number={},
  pages={01-08},
  doi={10.1109/ISSREW67781.2025.00059}}

@inproceedings{reimers-gurevych-2019-sentence,
  title = {Sentence-{BERT}: Sentence Embeddings using {S}iamese {BERT}-Networks},
  author = {Reimers, Nils and Gurevych, Iryna},
  editor = {Inui, Kentaro and Jiang, Jing and Ng, Vincent and Wan, Xiaojun},
  booktitle = {Proceedings of the 2019 Conference on Empirical Methods in Natural Language Processing and the 9th International Joint Conference on Natural Language Processing (EMNLP-IJCNLP)},
  month = nov,
  year = {2019},
  address = {Hong Kong, China},
  publisher = {Association for Computational Linguistics},
  url = {https://aclanthology.org/D19-1410/},
  doi = {10.18653/v1/D19-1410},
  pages = {3982--3992}
}

@inproceedings{wang-etal-2020-minilm,
  title = {MiniLM: Deep Self-Attention Distillation for Task-Agnostic Compression of Pre-Trained Transformers},
  author = {Wang, Wenhui and Wei, Furu and Dong, Li and Bao, Hangbo and Yang, Nan and Zhou, Ming},
  editor = {Larochelle, Hugo and Ranzato, Marc'Aurelio and Hadsell, Raia and Balcan, Maria-Florina and Lin, Hsuan-Tien},
  booktitle = {Advances in Neural Information Processing Systems 33},
  year = {2020},
  url = {https://proceedings.neurips.cc/paper_files/paper/2020/hash/3f5ee243547dee91fbd053c1c4a845aa-Abstract.html}
}

@misc{sentence-transformers-all-minilm-l6-v2,
  author = {{Sentence Transformers}},
  title = {{sentence-transformers/all-MiniLM-L6-v2}},
  howpublished = {\url{https://huggingface.co/sentence-transformers/all-MiniLM-L6-v2}},
  note = {Hugging Face model card, accessed April 22, 2026}
}

@INPROCEEDINGS{llmLogPreprocess2024,
  author={Ma, Zeyang and Chen, An Ran and Kim, Dong Jae and Chen, Tse-Hsun Peter and Wang, Shaowei},
  booktitle={2024 IEEE/ACM 46th International Conference on Software Engineering (ICSE)}, 
  title={LLMParser: An Exploratory Study on Using Large Language Models for Log Parsing}, 
  year={2024},
  volume={},
  number={},
  pages={1209-1221},
  doi={10.1145/3597503.3639150}}
  \end{document}